\documentclass[10pt,a4paper]{article}
\pdfoutput=1
\usepackage{jheppub}

\usepackage{hyperref}
\usepackage{color}
\usepackage{graphicx}
\usepackage{physics}
\usepackage{bm}
\usepackage{amssymb}
\usepackage{mathrsfs} 
\usepackage{xfrac} 
\usepackage{comment}

\newcommand{\ba}[1]{\begin{align} #1 \end{align}}
\newcommand{\bes}[1]{\begin{equation}\begin{split} #1 \end{split}\end{equation}}
\newcommand{\bsa}[2]{\begin{subequations}\label{#1}\begin{align} #2 \end{align}\end{subequations}}
\newcommand{\com}{\;,}
\newcommand{\per}{\;.}
\newcommand{\ie}[1]{i.e.}
\newcommand{\eg}[1]{e.g.}
\newcommand{\cf}[1]{cf.}

\newcommand{\nn}{\nonumber \\}

\newcommand{\del}[1]{\nabla_{\! \! #1}}
\newcommand{\metric}{{\fontfamily{times}\selectfont\textit{g}}}
\newcommand{\dete}{e}

\newcommand{\fref}[1]{Fig.~\ref{#1}}

\newcommand{\sref}[1]{Sec.~\ref{#1}}

\newcommand{\aref}[1]{Appendix~\ref{#1}}

\newcommand{\pref}[1]{(\ref{#1})}
\newcommand{\eref}[1]{Eq.~(\ref{#1})}

\newcommand{\rref}[1]{Ref.~\cite{#1}}
\newcommand{\rrefs}[1]{Refs.~\cite{#1}}

\newcommand{\Mpl}{M_\mathrm{Pl}}
\newcommand{\MPl}{\Mpl}
\newcommand{\MP}{\Mpl}
\newcommand{\ee}{\mathrm{e}}
\newcommand{\ii}{\mathrm{i}}
\newcommand{\xvec}{{\bm x}}

\newcommand{\kvec}{{\bm k}}

\newcommand{\pvec}{{\bm p}}

\newcommand{\SM}[1]{\text{\sc sm}}

\newcommand{\RH}[1]{{\text{\sc rh}}}
\newcommand{\RM}[1]{{\text{\sc rm}}}
\newcommand{\EQ}[1]{{\text{\sc eq}}}
\newcommand{\ssfrac}[2]{\sfrac{#1\mkern-2.2mu}{#2}}
\newcommand{\half}{\ssfrac{1}{2}}
\newcommand{\onehalf}{\ssfrac{1}{2}}
\newcommand{\threehalf}{\ssfrac{3}{2}}
\newcommand{\pprime}{{\prime\prime}}

\newcommand{\GeV}{\ \mathrm{GeV}}

\begin{document}

\title{\centering
Creation of spin-$\threehalf$ dark matter via cosmological gravitational particle production
}

\author[a]{Edward W. Kolb,}
\author[b]{Andrew J. Long,}
\author[c]{Evan McDonough,}
\author[a]{and Jingyuan Wang}

\affiliation[a]{Kavli Institute for Cosmological Physics and Enrico Fermi Institute, The University of Chicago, 5640 South Ellis Avenue, Chicago, IL 60637, U.S.A.}
\affiliation[b]{Department of Physics and Astronomy, Rice University, Houston, TX, 77005, U.S.A.}
\affiliation[c]{Department of Physics, University of Winnipeg, Winnipeg MB, R3B 2E9, Canada}

\abstract{
We study the cosmological gravitational particle production (CGPP) of spin-$\threehalf$ particles during and after cosmic inflation, and map the parameter space that can realize the observed dark matter density in stable spin-$\threehalf$ particles. Originally formulated by Rarita and Schwinger, the relativistic  theory of a massive spin-$\threehalf$ field later found a home in supergravity as the superpartner of the graviton, and in nuclear physics as baryonic resonances and nuclear isotopes. We study a minimal model realization, namely a free massive spin-$\threehalf$ field minimally coupled to gravity, and adopt the name raritron for this field. We demonstrate that CGPP of raritrons crucially depends on the hierarchy between the raritron mass $m_{3/2}$ and the Hubble parameter at the end of inflation $H_e$, with high-mass and low-mass cases  distinguished by the evolution of the sound speed $c_s$ of the longitudinal (helicity-$\onehalf$) mode, which is approximately unity at all times for heavy (relative to Hubble) raritrons and can become small or vanish for lighter raritrons, leading to a dramatic enhancement of production of high momentum particles in the latter case. Assuming the raritrons are stable, this leads to a wide parameter space to produce the observed dark matter density. Finally, we consider a time-dependent raritron mass, which can be chosen to remove the vanishing sound speed of the longitudinal mode, but which nonetheless enhances the production relative to the constant high-mass case, and in particular does not necessarily tame the high momentum tail of the spectrum.  We perform our calculations using the Bogoliubov formalism and compare, when applicable, to the Boltzmann formalism. }


\maketitle

\setlength{\parskip}{0.5ex}

\section{Introduction}
\label{sec:intro}

While the evidence for dark matter is overwhelming, there are but few clues to the fundamental properties, such as the mass and spin of the constituent dark matter particles. Indeed, the mass of dark matter could range over many orders of magnitude, and there is no constraint on the spin. One approach to progress in light of this uncertainty is to focus on production mechanisms for the observed dark matter relic density. Theoretical studies of production mechanisms play a complementary role to experimental searches for dark matter, with the former used to identify target regions of parameter space for the latter. Along these lines, a particularly compelling possibility is that dark matter could have been produced primordially through the dynamics of cosmic inflation via purely gravitational interaction. This approach, termed cosmological gravitational particle production (CGPP), has been studied for a wide variety of particle masses and spins (for a review, see \cite{Kolb:2023ydq}). 

In this paper we explore the possibility that dark matter is a spin-$\threehalf$ particle.  Originating in the work of Rarita and Schwinger \cite{Rarita:1941mf}, the Rarita-Schwinger (RS) field found a later home in supergravity as the gravitino, the superpartner to the graviton. In this context the mass of the RS field is generated dynamically by the super-Higgs mechanism, analogous to the conventional Higgs mechanism of particle physics. The original RS model is realized in supergravity via orthogonal nilpotent chiral superfields \cite{Carrasco:2015iij}, which derive from string theory as effective description of anti-D3 branes \cite{Kallosh:2016aep}. However, supergravity is not the only home for the RS field,  and spin-$\threehalf$ fields also arise as a composite particle like the spin-$\threehalf$ isotope of Xenon ($^{131}$Xe has spin-$\threehalf$ and is {\it stable}), or like a spin-$\threehalf$ baryonic resonance (which is unstable).

In this work we focus on the RS field: a free spin-$\threehalf$ field of mass $m_{3/2}$. We consider this model as a low-energy effective field theory, and remain agnostic to the UV completion, e.g., into supergravity. Following the convention of \rref{Garcia:2020hyo}, we call these particles {\it raritrons}.  Raritrons are produced gravitationally during inflation and at the end of inflation.  That is the main phenomenon that our work seeks to study. 

Gravitational production of spin-$\threehalf$ particles has been studied by several groups over the years~
\cite{Kallosh:1999jj,Kallosh:2000ve,Giudice:1999yt,Giudice:1999am,Bastero-Gil:2000lgf,Hasegawa:2017hgd,Kolb:2021xfn,Kolb:2021nob,Terada:2021rtp,Dudas:2021njv,Antoniadis:2021jtg}.  Those studies were partly motivated by the gravitino problem in theories of supergravity. In supergravity theories which reduce to the RS model, such as \cite{Carrasco:2015iij}, several studies by different groups all arrive at the same conclusion~\cite{Hasegawa:2017hgd,Kolb:2021xfn,Kolb:2021nob,Terada:2021rtp,Dudas:2021njv,Antoniadis:2021jtg}: if the mass of the spin-$\threehalf$ field is smaller than the inflationary Hubble scale (by a calculable order one factor), then the helicity-$\onehalf$ polarization mode of the spin-$\threehalf$ particle develops a vanishing sound speed at the end of inflation.  This  substantially increases the amount of particle production.  It has been called ``catastrophic particle production''~\cite{Kolb:2021xfn,Kolb:2021nob}, because the spectrum continues to rise toward higher momentum, leading to a large and cutoff-dependent prediction for the total amount of particle production. Conversely, if the raritron mass remains above the Hubble scale at the end of inflation, then catastrophic production is avoided and a more modest (but still enhanced) amount of particle production occurs. 

In this work we calculate the CGPP of massive spin-$\threehalf$ particles during inflation and at the end of inflation.  Assuming these particles are stable we calculate their cosmological energy fraction today, and compare with the measured dark matter energy fraction to assess whether CGPP is a viable production mechanism for raritron dark matter. To this end we consider three model classes:  high-mass raritron dark matter, low-mass raritron dark matter, and an evolving-mass raritron motivated by single chiral superfield models of supergravity \cite{Giudice:1999yt,Giudice:1999am}. The low- and high-mass cases are defined with respect to the Hubble parameter at the end of inflation, by  $m_{3/2} \gtrsim H_e$ and  $m_{3/2} \lesssim H_e$ respectively. In the high-mass case, there is no catastrophic production, and it is a straightforward application of CGPP to compute the particle production and resulting relic density. In the low-mass case, CGPP leads to catastrophic production, which yields a divergent spectrum. To regulate the divergence, we introduce a UV cutoff, and we discuss the dependence of our results on the value of this cutoff.  In both the low- and high-mass cases, we find a large region of parameter space, in reheating temperature $T_{\rm RH}$ and mass $m_{3/2}$, where the produced particles can constitute the observed dark matter.

We also consider an evolving-mass raritron model defined by a mass $m_{3/2}(t)$ that evolves with time such that the effective complex sound speed has constant magnitude $|c_s(t)| = 1$. We find that the evolving mass, though it removes the gradient instability, does not preclude catastrophic production. This surprising result motivates future studies of gravitino production in supergravity \cite{Ema:2016oxl}.  We note that while evolving-mass raritron dark matter was proposed in the context of SUGRA~\cite{Giudice:1999yt,Giudice:1999am}, a complete numerical calculation of its spectrum has not been done before. 

The theoretical framework for this work is CGPP \cite{Kolb:2023ydq}. We perform the CGPP calculation in two different ways.  First, we use the Bogoliubov formalism to express the spectrum in terms of the mode functions, which we calculate by solving the mode equations with numerical methods.  This method applies to all regimes of mass and momentum, but can be computationally intensive. Second, we use the Boltzmann formalism, which yields analytical expressions for the spectrum across a window of momenta.  We show that these two methods agree for the window of momenta where the Boltzmann formalism is valid.  However, for the parameters of interest, the peak of the spectrum derived with the Bogoliubov formalism falls outside the window. The Boltzmann formalism therefore underestimates the total amount of particle production.   

The main results of this paper are as follows:
\begin{itemize}
    \item We demonstrate that a free spin-$\threehalf$ field (a raritron), can be efficiently produced via CGPP. If the raritron is stable it can constitute the observed relic density of dark matter.
    \item We confirm the catastrophic production of low-mass raritrons observed previously in  \rrefs{Hasegawa:2017hgd,Kolb:2021xfn,Kolb:2021nob}. If the divergence is assumed to be resolved by UV completion, and if one models this by a UV cutoff in the effective theory, then the integrated relic density can again be compatible with the observed dark matter abundance.
    \item We demonstrate that a time-dependent mass $m_{3/2}(t)$ can remove the vanishing sound speed behind the catastrophic production.  However, this does not {\it a priori} cure the divergent particle production. If the raritron mass is decreasing and the raritron becomes light following inflation, then we find that raritrons can be produced with a rising spectrum in comoving momentum $p$ once the physical momentum $p/a(t)$ falls below the inflaton mass.      We expect that the divergent production in this case is regulated by the decay of the inflaton at reheating, rendering the status of catastrophic production highly dependent on the duration of reheating and hence the inflation model. 
\end{itemize}
The paper is organized as follows: In Sec.~\ref{sec:modeling} we introduce the raritron, including the cosmological evolution of the raritron sound speed, and review the basics of CGPP. In Sec.~\ref{sec:spectrum} we present results for the CGPP of raritrons, focusing on three models:  low-mass, high-mass, and evolving-mass raritrons. We find the raritrons can constitute the observed dark matter relic density for a wide range of raritron mass and reheat temperatures. We conclude in Sec.~\ref{sec:conc} with a summary and discussion of directions for future work.  In \aref{app:Boltzmann} we present some details of the Boltzmann calculation of the spectrum of produced raritrons.

\section{Modeling raritron dark matter}
\label{sec:modeling}

In this section we discuss our model for spin-$\threehalf$ raritron dark matter and the cosmological setting in which it may be produced through its gravitational interactions. 

\subsection{Modeling inflation and reheating}
\label{sub:inflation}

We assume a scalar inflaton field $\phi(x)$ with scalar potential $V(\phi) = \tfrac{1}{2} m_\phi^2 \phi^2$ with $m_\phi = 1.7 \times 10^{13} \GeV$.  We call this model Quadratic Inflation.  We denote the Hubble scale at 60 e-foldings before the end of inflation by $H_{60}$, and we denote the Hubble scale at the end of inflation by $H_e$.  Their values are $H_{60} \approx 1.1 \times 10^{14} \GeV$ and $H_e \approx 8.5 \times 10^{12} \GeV$.  

Quadratic Inflation is ruled out by CMB observations, particularly the upper limit on the tensor-to-scalar ratio inferred from the amplitude of B-mode polarization \cite{Planck:2018jri}.  Nevertheless this model provides a convenient benchmark for CGPP calculations that allows for comparison across different research groups.  
We expect similar results for other models of inflation with comparable values of $H_e$.  

Since the inflaton potential is quadratic in $\phi$, the system is effectively matter dominated during reheating (the period between the end of inflation and the start of radiation domination).  We assume that reheating is accomplished by the perturbative decay of inflaton particles into Standard Model particles.  We don't model these interactions, but rather treat the reheating temperature as a free parameter.  We denote the plasma temperature at the start of radiation domination by $T_\RH{}$, and the corresponding Hubble parameter $H_\RH{}$ is given by $3 \Mpl^2 H_\RH{}^2 = \tfrac{\pi^2}{30} g_{*,\RH{}} T_\RH{}^4$ where $\Mpl \approx 2.435 \times 10^{18} \GeV$.  We are generally interested in values of the reheating temperature in the range $100 \GeV < T_\RH{} < 10^{16} \GeV$ so that the full Standard Model is thermalized in the plasma and $g_{*,\RH{}} \approx 106.75$.  Energy conservation imposes $H_\RH{} \leq H_e$, which implies $T_\RH{} \leq T_\RH{}^\mathrm{max} = (2.5 \times 10^{15} \GeV) (H_e / 8.5 \times 10^{12} \GeV)^{1/2} (g_{*,\RH{}} / 106.75)^{-1/4}$. We focus on models for which $H_\RH{} < m_{3/2}$. For the evolving-mass raritron models, this condition is assumed to hold at all times.  When the Hubble parameter at the start of reheating is smaller than the mass of the particle experiencing gravitational production, it has been called ``late-reheating regime'' \cite{Kolb:2023ydq}.  

\subsection{Modeling spin-$\threehalf$ raritron}
\label{sub:raritron}

Spin-$\threehalf$ raritron particles are the quantum excitations of a spin-$\threehalf$ raritron field.  We model the raritron field and its interactions with gravity by the action~\cite{Freedman:2012zz} 
\ba{\label{eq:action}
    S[\Psi_\mu(x), e_\mu^a(x)] 
        = \int \! \dd^4 x \, \dete \, \Bigl[ 
    \tfrac{\ii}{4} \bar{\Psi}_\mu \gamma^{\mu\nu\rho} \del{\nu} \Psi_\rho 
    - \tfrac{\ii}{4} \bar{\Psi}_\mu \overleftarrow{\del{\nu}} \gamma^{\mu\nu\rho} \Psi_\rho 
    + \tfrac{1}{2} m_{3/2} \bar{\Psi}_\mu \gamma^{\mu\nu} \Psi_\nu 
    \Bigr] 
    \;,
}
which generalizes the Rarita-Schwinger action~\cite{Rarita:1941mf} to curved spacetime.  Here $\Psi_\mu(x)$ is the spin-$\threehalf$ raritron field, $m_{3/2}$ is its mass parameter, $e_\mu^a(x)$ is the veirbein field, $\metric_{\mu\nu}(x) = e_\mu^a(x) e_\nu^b(x) \eta_{ab}$ is the spacetime metric field, and $\eta_{ab}$ is the Minkowski spacetime metric.  Definitions for the other symbols can be found in \rrefs{Freedman:2012zz,Kolb:2023ydq}.  To study CGPP we set $\metric_{\mu\nu}(x) = \metric_{\mu\nu}^\mathrm{(FLRW)}(\eta,\xvec)$, which is discussed further in \sref{sub:Bogo}. The raritron field is required to be self-conjugate under the charge conjugation operation $\mathsf{C}$, similar to a spin-$\onehalf$ Majorana spinor field, and therefore it does not have distinct particle and antiparticle excitations. 

We will find it useful to distinguish three classes of models, which we call high-mass raritron, low-mass raritron, and evolving-mass raritron.  For high-mass and low-mass raritron models, the variable $m_{3/2}$ is a constant parameter, and the distinction between high- and low-mass involves a comparison with $H_e$ that we specify below.  For evolving-mass raritron models, the function $m_{3/2}(t)$ is given by a formula that we specify below. We note that particle production for the evolving-mass raritron models is not -- strictly speaking -- the usual CGPP, since the time-dependent mass is not a consequence of the cosmological expansion.

We assume that the raritron field only interacts with gravity in order to demonstrate that CGPP can explain the origin of raritron dark matter, even in the absence of non-gravitational interactions.  More specifically, we assume a minimal coupling to gravity, as specified by the action in \eref{eq:action}.  However, nonminimal couplings to gravity have been studied for fields of spin-0 and spin-1, leading to novel phenomena \cite{Kolb:2022eyn,Garcia:2023qab,Capanelli:2024rlk}. One may want to consider other operators that couple the raritron field with gravity.  A simple example is $\Delta S \sim R\, \bar{\Psi}_\mu \gamma^{\mu\nu} \Psi_\nu$, which replaces the raritron mass parameter $m_{3/2}$ with the Ricci scalar field $R(x)$.  Since these interactions are operators of mass dimension $> 4$, it is reasonable to expect that they would be suppressed by the UV cutoff $\Lambda$ of the effective field theory.  This cutoff may be as high as the Planck scale $\Lambda \sim \Mpl$ or the scale of supersymmetry breaking.  Provided that the cutoff is above the scale of inflation $\Lambda \gg H_e$, we expect that the effect of these interactions on CGPP can be neglected.  

\subsection{Raritron stability}
\label{sub:stability}

If the raritron should provide a candidate for the cold dark matter, it must be a stable particle or at least cosmologically long lived.  If the raritron only interacts with gravity, as we have assumed at \eref{eq:action}, then it is stable as a consequence of fermion-number conservation:  a raritron cannot decay into gravitons alone.  If the raritron is allowed to interact with the SM particle content, this may open channels for it to decay.  Such interactions are expected to arise for a spin-$\threehalf$ gravitino in theories of SUGRA~\cite{Krauss:1983ik,Ellis:1984eq}. Depending on the nature of these interactions, it may be possible to calculate the raritron's lifetime in an identical fashion to the gravitino's lifetime.  This topic has been studied extensively, see for example \rrefs{Krauss:1983ik,Ellis:1984eq}, since late-decaying gravitinos are one aspect of the Gravitino Problem.  By drawing a parallel with the gravitino, which couples with gravitational-strength interactions, we estimate the raritron's lifetime to be 
\begin{align}
    \tau_{3/2} 
    & = \biggl( \frac{N_\mathrm{ch}}{2\pi} \frac{m_{3/2}^3}{\MP^2} \biggr)^{-1} 
    \approx (78 \, \mathrm{Gyr}) \biggl( \frac{N_\mathrm{ch}}{10} \biggr)^{-1} \biggl( \frac{m_{3/2}}{10 \, \mathrm{MeV}} \biggr)^{-3} 
\end{align}
where $N_\mathrm{ch}$ is the number of kinematically accessible channels and $\MP \approx 2.43 \times 10^{18} \, \mathrm{GeV}$ is the reduced Planck mass.  Therefore, if there are channels available for the raritron to decay via gravitational-strength interactions, then the raritron mass would have to be $m_{3/2} \lesssim 10 \, \mathrm{MeV}$ to ensure a cosmologically stable raritron ($\tau_{3/2} \gtrsim 13.7 \, \mathrm{Gyr}$).  If the raritron's decay products include SM particles, then decays of raritron dark matter would produce highly boosted cosmic rays that are not observed, implying a much stronger lower bound on $\tau_{3/2}$.  For the remainder of this article we focus our attention on the minimal model in \eref{eq:action}, which features stable raritrons. 

\subsection{Bogoliubov formalism for CGPP}
\label{sub:Bogo}

In this subsection we discuss how the phenomenon of CGPP may be studied using the Bogoliubov formalism~\cite{Kolb:2023ydq}.  CGPP occurs when the cosmological expansion induces explicit time dependence to the equations of motion for spectator fields.  We model the spacetime during inflation and at the end of inflation as the homogeneous and isotropic FLRW spacetime.  The spacetime metric $\metric_{\mu\nu}$ is used to calculate the invariant interval, which is $(\dd s)^2 = \metric_{\mu\nu} \dd x^\mu \dd x^\nu = (\dd t)^2 - a(t)^2 |\dd \xvec|^2 = a(\eta)^2 [ (\dd \eta)^2 - |\dd \xvec|^2 ]$ where $t$ is cosmic time, $\eta$ is conformal time, $\xvec$ is the comoving spatial coordinate 3-vector, and $a$ is the scale factor. 

In the FRW spacetime, the Rarita-Schwinger field equation yields equations of motion for the helicity-$\onehalf$ and helicity-$\threehalf$ polarization modes.  In the Dirac representation of the gamma matrices, these equations are~\cite{Hasegawa:2017hgd,Kolb:2021xfn} 
\bsa{eq:mode_eqns}{
    \ii \frac{\dd}{\dd\eta} \begin{pmatrix} u_{A,k,\threehalf} \\ u_{B,k,\threehalf} \end{pmatrix} 
    & = \begin{pmatrix} a m_{3/2} & k \\ k & - a m_{3/2} \end{pmatrix} \begin{pmatrix} u_{A,k,\threehalf} \\ u_{B,k,\threehalf} \end{pmatrix} \\ 
    \ii \frac{\dd}{\dd\eta} \begin{pmatrix} u_{A,k,\onehalf} \\ u_{B,k,\onehalf} \end{pmatrix} 
    & = \begin{pmatrix} a m_{3/2} & c_s k \\ c_s^\ast k & - a m_{3/2} \end{pmatrix} \begin{pmatrix} u_{A,k,\onehalf} \\ u_{B,k,\onehalf} \end{pmatrix} 
    \;.
}
Here $\kvec$ is the comoving wavevector, $k = |\kvec|$ is the corresponding comoving wavenumber, and the mode functions are denoted by $u_{A,k,\threehalf}(\eta)$ and so on.  The equations of motion for the helicity-$\threehalf$ polarization modes are identical to the equations of motion for the helicity-$\onehalf$ polarization modes of a spin-$\onehalf$ Dirac or Majorana spinor field~\cite{Chung:2011ck}.  The equations of motion for the helicity-$\onehalf$ polarization modes have an additional time-dependent factor \cite{Kolb:2021xfn}:
\ba{\label{eq:cs}
    c_s(\eta) & = \frac{m_{3/2}^2 - \dfrac{1}{3} H^2 - \dfrac{1}{9} R + \dfrac{2 \ii}{3} \dfrac{1}{a} \dfrac{\dd}{\dd \eta} m_{3/2}}{(m_{3/2} - \ii H)^2} 
    \;,
}
which can be interpreted as a complex sound speed \cite{Hasegawa:2017hgd}.  
Note that the Hubble parameter is $H(\eta) = a^\prime / a^2$, and the FLRW Ricci scalar is $R(\eta) = - 6 a^\pprime / a^3$.  Here we have allowed for the possibility that the raritron mass $m_{3/2}$ may vary in time.  In the next subsection we discuss the implications of a time-dependent sound speed.  By performing a change of variables, the mode equations \pref{eq:mode_eqns} can be expressed equivalently as~\cite{Hashiba:2022bzi} 
\bsa{eq:mode_eqns_alpha_beta}{
    \ii \frac{\dd}{\dd\eta} \begin{pmatrix} \tilde{\alpha}_{k,\threehalf} \\ \tilde{\beta}_{k,\threehalf} \end{pmatrix} 
    & = \begin{pmatrix} 0 & -\ii \mu_{\threehalf} \, \ee^{2 \ii \Phi_{\threehalf}} \\ \ii \mu_{\threehalf}^\ast \, \ee^{-2 \ii \Phi_{\threehalf}} & 0 \end{pmatrix} \begin{pmatrix} \tilde{\alpha}_{k,\threehalf} \\ \tilde{\beta}_{k,\threehalf} \end{pmatrix} \\
    \ii \frac{\dd}{\dd\eta} \begin{pmatrix} \tilde{\alpha}_{k,\onehalf} \\ \tilde{\beta}_{k,\onehalf} \end{pmatrix} 
    & = \begin{pmatrix} 0 & -\ii \mu_{\onehalf} \, \ee^{2 \ii \Phi_{\onehalf}} \\ \ii \mu_{\onehalf}^\ast \, \ee^{-2 \ii \Phi_{\onehalf}} & 0 \end{pmatrix} \begin{pmatrix} \tilde{\alpha}_{k,\onehalf} \\ \tilde{\beta}_{k,\onehalf} \end{pmatrix} 
    \;,
}
where 
\begin{subequations}\label{eq:Fourier}
\begin{align}
    \omega_{\threehalf}(\eta) & = \sqrt{k^2 + a^2 m_{3/2}^2} 
    \ , \qquad 
    \omega_{\onehalf}(\eta) = \sqrt{|c_s|^2 k^2 + a^2 m_{3/2}^2} \ , \\
    \Phi_{\threehalf}(\eta) & = \int_{-\infty}^\eta \! \dd\eta^\prime \, \omega_{\threehalf}(\eta^\prime) 
    \ , \qquad 
    \Phi_{\onehalf}(\eta) = \int_{-\infty}^\eta \! \dd\eta^\prime \, \omega_{\onehalf}(\eta^\prime) \ ,  \\
    \mu_{\threehalf}(\eta) & = \frac{k a^2 H m_{3/2} + k a \partial_\eta m_{3/2}}{2 \omega_{\threehalf}^2} 
    \ , \\ 
    \mu_{\onehalf}(\eta) & = \frac{|c_s| k a^2 H m_{3/2} + |c_s| k a \partial_\eta m_{3/2} - a m_{3/2} \partial_\eta |c_s| k - \ii |c_s| k \omega_{\onehalf} \partial_\eta \mathrm{arg}[c_s]}{2 \omega_{\onehalf}^2} \, \ee^{2 \ii \delta} \ ,  \\
    \delta(\eta) & = \int_{-\infty}^\eta \! \dd\eta^\prime \, \frac{a(\eta^\prime) m_{3/2}(\eta^\prime)}{2 \omega_{\onehalf}(\eta^\prime)} \, \partial_{\eta^\prime} \mathrm{arg}[c_s(\eta^\prime)] 
    \ .
\end{align}
\end{subequations}
If CGPP is inefficient then $\tilde{\alpha} \approx 1$ and
\ba{\label{eq:FT}
    \tilde{\beta}_{k,a}(\eta) 
    & \approx \int_{-\infty}^{\eta} \! \dd\eta^\prime \, \mu_a^\ast(\eta^\prime) \, \ee^{-2 \ii \Phi_a(\eta^\prime)} 
    = \int_{-\infty}^{\bar{\theta}(\eta)} \! \dd\theta^\prime \, \frac{\mu_a^\ast(\bar{\eta}(\theta^\prime))}{\omega_a(\bar{\eta}(\theta^\prime))} \, \ee^{-2 \ii \theta^\prime}
    \com
}
for $a = \threehalf, \onehalf$.  In the second equality we changed variables from $\eta$ to $\theta$ by noting that $\omega_a(\eta)$ is strictly positive, and by using $\dd\theta = \dd\eta \, \omega_a(\eta)$ with the initial condition $\theta=-\infty$ for $\eta=-\infty$, which has solutions $\eta = \bar{\eta}(\theta)$ and $\theta = \bar{\theta}(\eta)$.  For $\eta,\theta \to \infty$ the integral is the Fourier transform of $\mu_a^\ast / \omega_a$ evaluated at $2$. 

The CGPP calculation proceeds as follows; see \rref{Kolb:2021xfn} for additional details.  First, one specifies an initial condition for the mode functions.  We impose the Bunch-Davies condition $\tilde{\alpha}(-\infty) = 1$ and $\tilde{\beta}(-\infty) = 0$, which corresponds to having only the positive-frequency oscillation mode $\propto \ee^{-\ii k \eta}$ present at asymptotically early time when a mode with comoving wavenumber $k$ is deep inside the horizon during inflation.  Next, one calculates the time-dependent mode functions, which solve the mode equations subject to this initial condition.  We employ numerical integration methods, since exact analytical solutions are not available for the time-dependent backgrounds that we consider. Next, one calculates the Bogoliubov coefficients $\beta_{k,\onehalf}$ and $\beta_{k,\threehalf}$ for each Fourier mode and both polarization modes.  The Bogoliubov coefficient $\beta_{k,a} = \lim_{\eta\to\infty} u_{B,k,a}(\eta) = \lim_{\eta\to\infty} \tilde{\beta}_{k,a}(\eta)$ represents the complex amplitude of the negative-frequency oscillation mode at late time for a solution of the mode equations that was initialized in the positive-frequency oscillation mode at early time.  For this system the definition of the Bogoliubov coefficient imposes $|\beta_{k,a}| \leq 1$, which is a manifestation of Pauli-blocking, since spin-$\threehalf$ particles are fermions. Next, the comoving number density spectrum (per logarithmically spaced comoving momentum interval) is calculated as 
\bes{\label{eq:a3np_from_beta2}
    a^3 n_{p,\onehalf} = 2 \frac{p^3}{2\pi^2} |\beta_{p,\threehalf}|^2 
    \qquad \text{and} \qquad 
    a^3 n_{p,\threehalf} = 2 \frac{p^3}{2\pi^2} |\beta_{p,\threehalf}|^2
    \;,
}
where we set wavenumber $k = p$.  The factor of $2$ accounts for the positive and negative helicity in each polarization mode. The comoving number density is calculated as $a^3 n = a^3 n_{\threehalf} + a^3 n_{\onehalf}$ where $a^3 n_{\onehalf} = \int_0^\infty \! \tfrac{\dd p}{p} \, a^3 n_{p,\threehalf}$ and $a^3 n_{\onehalf} = \int_0^\infty \! \tfrac{\dd p}{p} \, a^3 n_{p,\threehalf}$.  Finally, assuming that the raritron is cosmologically long lived and nonrelativistic before today, its cosmological energy fraction is calculated: 
\begin{align}\label{eq:Omegah2}
    \Omega h^2 \approx \bigl( 1.2 \times 10^4 \bigr)\, \biggl( \frac{m_{3/2}}{H_e} \biggr) \biggl( \frac{H_e}{10^{12} \, \mathrm{GeV}} \biggr)^{\!\!2} \biggl( \frac{T_\RH{}}{10^9 \, \mathrm{GeV}} \biggr) \biggl( \frac{a^3 n_{\threehalf} + a^3 n_{\onehalf}}{a_e^3 H_e^3} \biggr) 
    \per 
\end{align}
For reference, the observed dark matter abundance corresponds to $\Omega h^2 \approx 0.12$.  

\subsection{Sound speed evolution}
\label{sub:sound_speed}

As one can see from \eref{eq:mode_eqns}, a peculiar feature of CGPP for spin-$\threehalf$ particles is the time-dependent sound speed $c_s(\eta)$ of the helicity-$\onehalf$ polarization modes \eqref{eq:cs}. For the purposes of studying CGPP we find it convenient to classify raritron models based on how the sound speed behaves during inflation and soon after inflation is ended.  If the raritron mass $m_{3/2}$ is always much larger than $H(\eta)$ and $R(\eta)$, then the sound speed $c_s(\eta) \approx 1$ is approximately static and equal to unity.  For smaller values of the raritron mass, it is possible that $H^2/3 + R/9$ momentarily passes through $m_{3/2}^2$, and so $c_s(\eta)$ passes through zero.  This crossing may occur just one time or possibly at many times, $\eta_n = \eta_1$, $\eta_2$, $\cdots$, while the inflaton field oscillates after inflation.  Alternatively, if the raritron mass evolves with time, then its evolution may ensure that $|c_s(\eta)|=1$ at all times.  

\begin{figure}[t]
    \centering
    \includegraphics[width=1\linewidth]{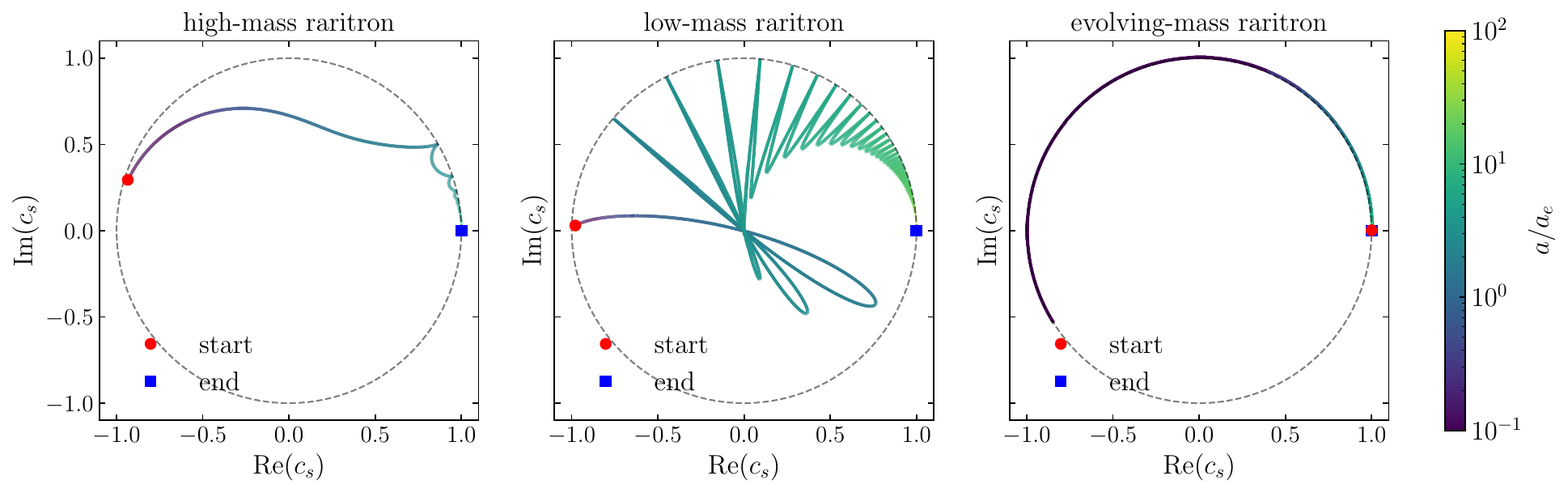} \hfill 
    \caption{\label{fig:sound_speed}
    Evolution of the complex sound speed $c_s(\eta)$.  \textit{Left:}  High-mass raritron with $m_{3/2}/H_e = 1.0$.  \textit{Middle:}  Low-mass raritron with $m_{3/2}/H_e = 0.10$.  \textit{Right:} Evolving-mass raritron with initial condition $m_{3/2}/H_e = 6 \times 10^3$.  
    Note that for the evolving-mass raritron both the initial and final values are $\mathrm{Re}(c_s)=1,\ \mathrm{Im}(c_s)=0.$
    }
\end{figure}

With these behaviors in mind, we identify three raritron model classes: 
\begin{subequations}\label{eq:raritron_models}
\begin{align}
    \text{high-mass raritron:} & \qquad 
    0 < |c_s(\eta)| \leq 1 \quad \text{at all times} \\ 
    \text{low-mass raritron:} & \qquad 
    |c_s(\eta_n)| = 0 \quad \text{at one or several times $\eta_n$} \\ 
    \text{evolving-mass raritron:} & \qquad 
    |c_s(\eta)| = 1 \quad \text{at all times} 
    \;.
\end{align}
\end{subequations}
For the high-mass and low-mass raritron models, we treat $m_{3/2}$ as a constant parameter and set $\partial_\eta m_{3/2} = 0$ in \eref{eq:cs}.  Since $c_s(\eta)$ depends upon $H(\eta)$ and $R(\eta)$, the dividing line between high-mass and low-mass raritron models is tied up with the model of inflation.  For Quadratic Inflation we can use Fig.~3 of \rref{Kolb:2021xfn} to identify $m_{3/2} / H_e \gtrsim 0.4$ as high-mass raritron models and $m_{3/2} / H_e \lesssim 0.4$ as low-mass raritron models.  When we study evolving-mass raritron models, we treat the raritron mass as a time-dependent variable that evolves such that the complex sound speed maintains a constant, unit magnitude~\cite{Giudice:1999yt}  
\begin{align}\label{eq:cs_condition}
    |c_s(\eta)|^2 = 1 
    \qquad \Rightarrow \qquad  
    \bigl( a^{-1}\partial_\eta m_{3/2} \bigr)^2 = \bigl( 3 m_{3/2}^2 + H^2 - \tfrac{1}{6} R \bigr) \bigl( 2 H^2 + \tfrac{1}{6} R \bigr) 
    \;.
\end{align}
If $m_{3/2}(\eta)$ satisfies this differential equation then the sound speed reduces to a time-dependent phase, $c_s(\eta) = \mathrm{exp}[\ii \zeta(\eta)]$.  Examples of the evolution of $c_s(\eta)$ are illustrated in \fref{fig:sound_speed}. Note that the high-mass raritron maintains $0 < |c_s(\eta)| \leq 1$ at all times, the low-mass raritron has numerous zero crossings where $c_s(\eta_n) = 0$, and the evolving-mass raritron maintains $|c_s(\eta)|=1$ at all times.  For all three calculations the late-time behavior is $c_s \to 1$ as $\eta \to \infty$, which can be understood from \eref{eq:cs} by noting that $m_{3/2}^2 \gg H^2, R$ at late times. 

While we remain agnostic as to the UV completion of the raritron, the embedding into supergravity provides one example of an evolving mass. Take for example ${\cal N}=1$ supergravity with a single chiral superfield $\Phi$, a  superpotential $W(\Phi)$, and a K\"{a}hler potential $K(\Phi,\bar{\Phi})$. The gravitino mass is field-dependent and is given by
\begin{equation} 
    m_{3/2} = \ee^{K(\Phi,\Bar{\Phi})/2 M_{\rm pl}^2}\frac{W(\Phi)}{M_{\rm pl}^2}.
\end{equation}
If $\Phi$ is time-dependent, this generates a time-dependence of the gravitino mass. For example, the model $
    K = \Phi \bar{\Phi}$ , $ W(\Phi) = m_\phi \Phi^2 /2 $ leads to
 \begin{align}\label{eq:m321}
	m_{3/2} = \ee^{K(\Phi,\bar{\Phi})/2\Mpl^2} \frac{W(\Phi)}{\Mpl^2} = \ee^{\phi^2/4\Mpl^2} \, \frac{m_\phi}{4\Mpl^2} \, \phi^2 \com
\end{align}
where we define $\Phi=\tfrac{1}{\sqrt{2}}\, \phi\, \ee^{\ii\sigma/\Mpl}$ and set $\sigma=0$. One may easily check (see Ref.~\cite{Kolb:2021xfn}) that this satisfies the condition Eq.~\eqref{eq:cs_condition} and hence $|c_s|=1$.  (Though we note this simple model does not provide a model of inflation, since $V(\phi)$ is too steep, and since supersymmetry is restored in the minimum of $V(\phi)$).

This easily generalizes to multiple chiral superfields  $\Phi^I$. The time-dependence of the gravitino mass is given in this case by \cite{Kolb:2021xfn}
\begin{equation}
\Mpl^4 \left|\,\dot{m}_{3/2}\right|^2  = \ee^{K(\Phi,\Bar{\Phi})/\Mpl^2} \left| \sum_I   \dot{\Phi}^I  D_I W(\Phi) \right|^2 \com
\end{equation}
where $D_I \equiv \partial_I + \Mpl^{-2}\,K_{,I}$. From this sound speed is given by
\begin{align}
c_s^2 = 1 - 4  \frac{ (\vec{\dot{\Phi}}  \cdot \vec{\dot{\Phi}} ) (\vec{F} \cdot \vec{F} ) - (\vec{\dot{\Phi}} \cdot \vec{F})^2 }{(\vec{\dot{\Phi}} \cdot \vec{\dot{\Phi}} +  \vec{F} \cdot \vec{F})^2}.
\end{align}
where $F_{I}\equiv D_{I}W$. From this one may appreciate that the departure of $c_s$ from unity in supergravity is due to a misalignment of the field velocity $\dot{\Phi}^I$ and supersymmetry breaking $D_IW$. In the case of single chiral superfield, there is only field space direction and therefore no possibility of misalignment. The sound speed is unity at all times in models of a single chiral superfield.

In lieu of a concrete supergravity model, we may instead adopt Eq.~\eqref{eq:cs_condition} as the defining property of our evolving-mass raritron model. We do so in what follows. 

\subsection{Boltzmann formalism for CGPP}
\label{sub:Boltz}

Although the Bogoluibov formalism provides a complete description of CGPP across all scales (both inside and outside the horizon) and at all times (both during and after inflation), the calculations are not always analytically tractable and often numerically difficult at high wavenumber.  Therefore it's generally illuminating to appeal to a second description of CGPP, which is known as the Boltzmann formalism, where many calculations can be performed analytically.  The Boltzmann formalism describes CGPP at the end of inflation as arising from annihilations of inflaton particles through an $s$-channel graviton exchange~\cite{Ema:2018ucl,Kaneta:2022gug,Chakraborty:2025zgx}.  Provided that the raritron mass is below the inflaton mass, $m_{3/2} < m_\phi$, then 2-to-2 annihilations at time $\eta$ of inflaton particles with energy $E_\phi \approx m_\phi$ produce lighter raritron particles with comoving momentum $p \approx a(\eta) m_\phi$.  Annihilations begin at time $\eta_\ast$ soon after the end of inflation, when the inflaton field can be interpreted as a collection of cold particles; they end around the start of radiation domination, when the inflaton field is depleted by its decays into the thermal bath.  Consequently, annihilations populate the raritron spectrum across the range of comoving momenta, $a_* m_\phi \lesssim p \lesssim a_\RH{} m_\phi$.\footnote{The value of $a_*$ would depend on the value of $a$ at the onset of oscillations.  For quadratic inflation it is approximately $a_*=2a_e$, so the Boltzmann calculation would apply for $p/a_eH_e\gtrsim 4\sqrt{1-r^2}$. This is the choice made in Figs.~\ref{fig:highmass_spectra} and \ref{fig:lowmass_spectra}.} For 2-to-2 annihilations, the spectrum develops a power-law scaling $a^3 n_p \propto p^{-3/2}$~\cite{Ema:2018ucl}, because higher-$p$ modes are produced later when the inflaton density has reduced through redshifting.  For models with a constant raritron mass $m_{3/2} < m_\phi$ and sound speed $c_s = 1$, the authors of \rref{Kaneta:2023uwi} studied the production of raritrons through 2-to-2 inflaton annihilations. Using their matrix elements, we calculate the resultant raritron spectrum in \aref{app:Boltzmann}, and our results are 
\begin{subequations}\label{eq:a3np_Boltzmann}
\begin{align}
    \frac{a^3n_{p,\threehalf}}{a_e^3H_e^3} & = \frac{9}{256\pi}\frac{m_{3/2}^2}{m_\phi^2} \left( 1 - \frac{m_{3/2}^2}{m_\phi^2} \right)^{\!9/4}\left(\frac{a_eH_e}{p}\right)^{3/2}  \nonumber \\
    & \times\Theta(m_\phi - m_{3/2}) \, 
    \Theta\bigl( \sqrt{1-r^2} \, a_\ast m_\phi < p < \sqrt{1-r^2} \, a_\RH{} m_\phi \bigr)\\
    \frac{a^3n_{p,\half}}{a_e^3H_e^3} &  =\frac{1}{16\pi} \frac{m_\phi^2}{m_{3/2}^2} \left( 1 - \frac{m_{3/2}^2}{m_\phi^2} \right)^{\!9/4} \left(1-\frac{3}{2}\frac{m_{3/2}^2}{m_\phi^2} + \frac{9}{16}\frac{m_{3/2}^4}{m_\phi^4}\right)\left(\frac{a_eH_e}{p}\right)^{3/2} \nonumber \\
    & \times\Theta(m_\phi - m_{3/2}) \, 
    \Theta\bigl( \sqrt{1-r^2} \, a_\ast m_\phi < p < \sqrt{1-r^2} \, a_\RH{} m_\phi \bigr)
    \;.
    \end{align}
\end{subequations}
This power law is valid across a range of comoving momenta, which is approximately $4 a_e H_e \lesssim p \lesssim a_\RH{} H_e$, where the factor of $4$ accounts for the delay between the end of inflaton and the onset of inflaton oscillations. It is straightforward to generalize these results to allow for a constant sound speed $|c_s| \neq 1$, and we do so in \aref{app:Boltzmann}.  

For lower-spin fields, the $p^{-3/2}$ scaling can also be derived analytically from the Bogoliubov formalism~\cite{Chung:2018ayg,Kaneta:2022gug,Basso:2022tpd}.  Similarly, for lower-spin fields the generalization for many-to-2 annihilations has also been worked out~\cite{Basso:2022tpd}; although these channels are subleading to 2-to-2, they become the dominant channel if the 2-to-2 channel is kinematically blocked because $m_{3/2} > m_\phi$.  

It is important to emphasize that the Boltzmann formalism only captures CGPP happening at the end of inflation while the inflaton field is oscillating on its quadratic potential and can be viewed as a collection of cold particles~\cite{Chung:2018ayg,Kaneta:2022gug}.  In other words, it only allows one to calculate the spectrum at $p \gtrsim a_e m_\phi$.  One can derive a conservative limit on dark matter over-production by assuming that the spectrum vanishes for $p \lesssim a_e m_\phi$.  However, if one seeks to make a prediction for the present-day abundance of raritron dark matter, then calculating only the spectrum for $p \gtrsim 4 a_e m_\phi$ will lead to an underestimate that could be very large and dependent upon the model parameters.  In the next section we evaluate CGPP using both the Bogoliubov and Boltzmann formalisms.  

\section{Spectrum and abundance of gravitationally-produced raritrons}
\label{sec:spectrum}

In this section we present our main results for the spectrum and total abundance of raritrons that are produced gravitationally during inflation and at the end of inflation.  We consider each of the three raritron model classes in turn: high-mass raritron, low-mass raritron, and evolving-mass raritron. 

\subsection{High-mass raritron}
\label{sub:high_mass}

Here we study the class of models with a high-mass raritron.  As we noted already at \eref{eq:raritron_models}, the high-mass raritron maintains $0 < |c_s(\eta)| \leq 1$ at all times, and for Quadratic Inflation this condition corresponds to a large value for the raritron mass parameter $m_{3/2} \gtrsim 0.4 H_e$.  In the remainder of this subsection, first we'll discuss the spectrum of the helicity-$\threehalf$ polarization modes, then the helicity-$\onehalf$ polarization modes, and then the relic abundance constraints across parameter space. 

\begin{figure}[t]
    \centering
    \includegraphics[width=0.48\linewidth]{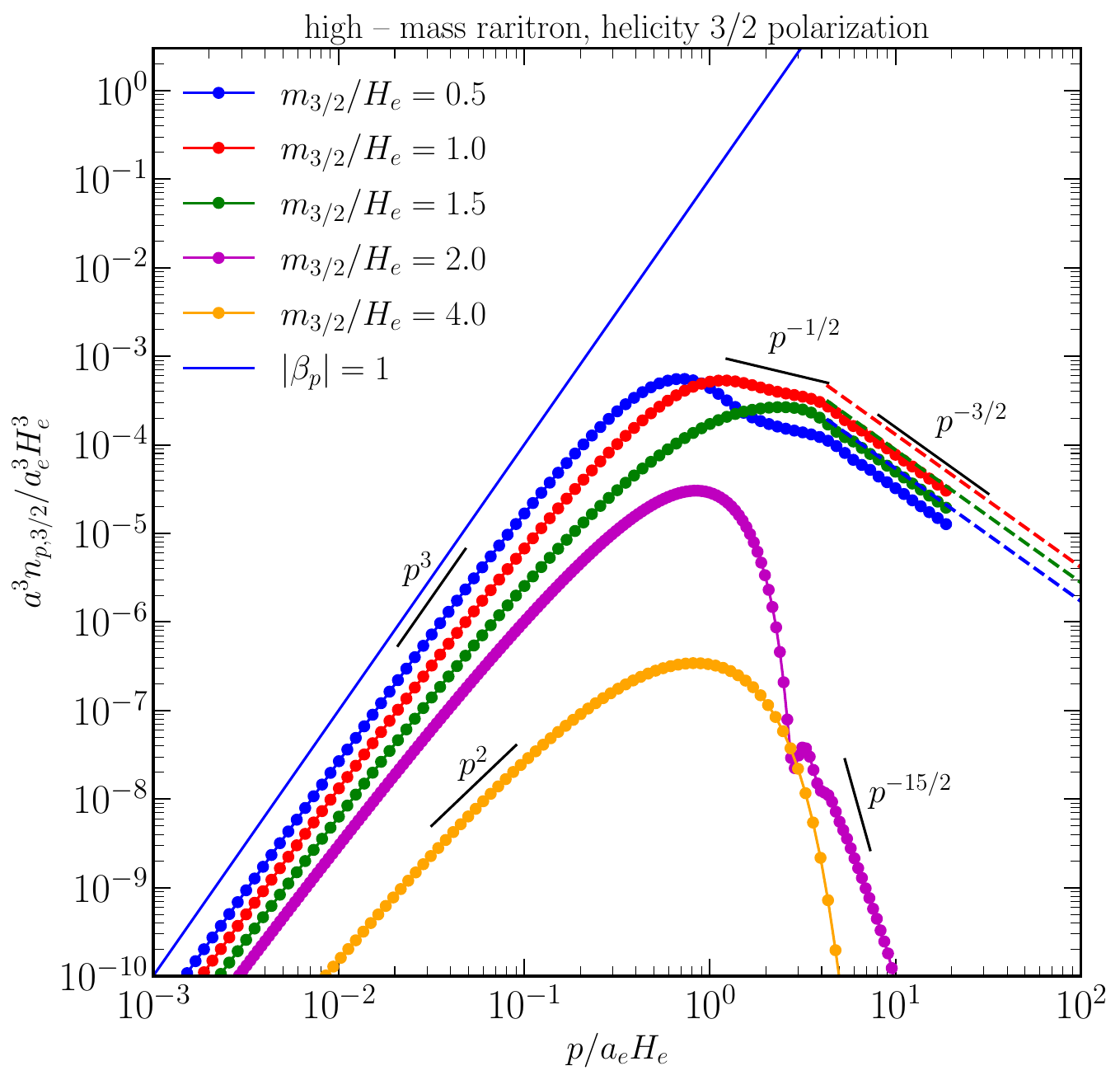}
    \hfill 
    \includegraphics[width=0.48\linewidth]{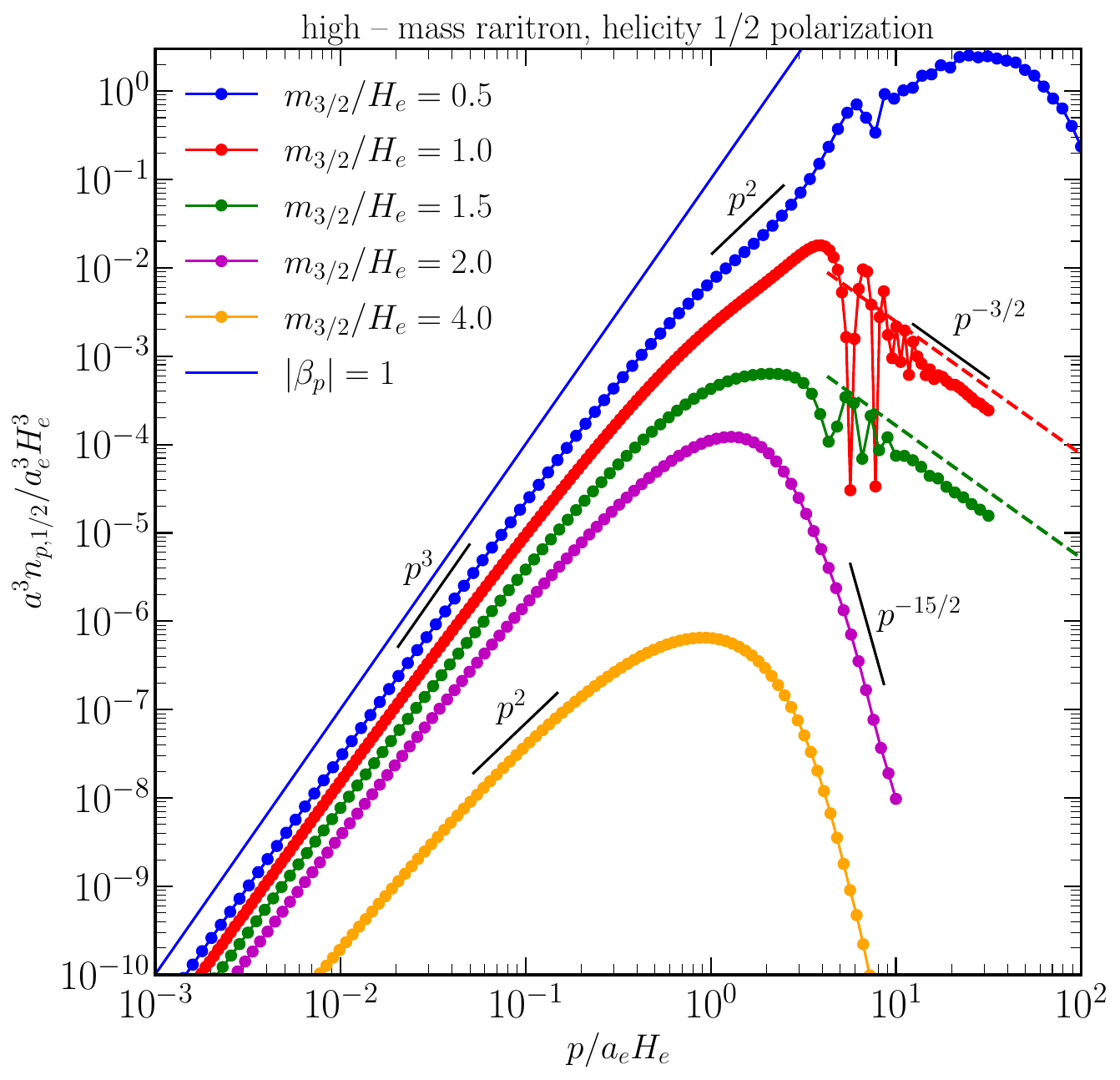}
    \caption{\label{fig:highmass_spectra}
    Spectra of CGPP for high-mass raritron models.  We plot the comoving number density spectra $a^3 n_p = a^3 \dd n / \dd \ln p$ as a function of the comoving momentum $p$ for several values of the raritron mass $m_{3/2}$.  \textit{Left:}  Helicity-$\threehalf$ polarization mode.  \textit{Right:}  Helicity-$\onehalf$ polarization mode.  The dashed lines show the analytical Boltzmann calculation for 2-to-2 scattering \pref{eq:a3np_Boltzmann}.  Pauli blocking prohibits the spectrum from rising above the blue diagonal where the Bogoliubov coefficient is maximal, $|\beta_p| = 1$. 
    }
\end{figure}

Now we consider the spectrum of gravitationally produced raritrons in the helicity-$\threehalf$ polarization modes.  The results of our numerical calculation are presented in the left panel of \fref{fig:highmass_spectra}.  For the curves labeled $m_{3/2} = 0.5$, $1.0$, and $1.5 H_e$, the spectrum rises as $p^3$, reaches a maximum at $p \approx a_e H_e$, has an intermediate behavior that falls as $p^{-1/2}$, and continues to fall as $p^{-3/2}$ above $p \approx 4 a_e H_e$ until $p = 20 a_e H_e$.  We are unable to calculate the spectrum reliably using numerical methods for larger values of $p$.  For the curve labeled $m_{3/2} = 2.0 H_e$, the spectrum begins rising as $p^3$, reaches a maximum at $p \approx a_e H_e$, falls as an exponential, and continues to fall as $p^{-15/2}$ above $p \approx 3 a_e H_e$ until $p \approx 10 a_e H_e$.  For the curve labeled $m = 4.0 H_e$, the spectrum rises as $p^2$, reaches a maximum at $p \approx a_e H_e$, and falls exponentially. The curve labeled $|\beta_p|=1$ indicates the maximum possible value of the spectrum at each comoving momentum, $a^3 n_{p,\threehalf} = p^3 / \pi^2$ for $|\beta_{p,\threehalf}| = 1$, since Pauli blocking prevents the Bogoliubov coefficient from rising above one. 

The spectrum of the helicity-$\threehalf$ polarization modes can be understood as follows.  The low-momentum power-law $\propto p^3$ is a consequence of the raritron's fermionic nature.  As noted previously in \rref{Chung:2011ck} in the context of CGPP for spin-$\onehalf$ particles, Pauli blocking leads to $|\beta_p|^2 \approx 1/2$ and the spectrum is $a^3 n_p = 2 (p^3 / 2\pi^2) |\beta_p|^2 \approx p^3 / 2\pi^2$.  The same arguments apply here, since the helicity-$\threehalf$ polarization modes of a spin-$\threehalf$ field obey the same equations of motion as a spin-$\onehalf$ field.  The high-momentum power-laws $\propto p^{-3/2}, p^{-15/2}$ can be explained by the Boltzmann formalism.  For $m_{3/2} < m_\phi \approx 2 H_e$, the annihilation of pairs of inflaton particles into pairs of raritron particles via an $s$-channel graviton exchange leads to the $p^{-3/2}$ scaling; this has been noted previously for spin-$\onehalf$ \cite{Ema:2019yrd} and spin-$\threehalf$ particles \cite{Kaneta:2023uwi}.  On the figure we also plot the analytical expressions in \eref{eq:a3np_Boltzmann} for comparison, which agree well with the $p^{-3/2}$ power law branch of the spectrum.  In fact, the Boltzmann formalism does a reasonably good job of capturing the bulk of the particle production (i.e., area under the curve), since it works well for $p \gtrsim 2 a_e H_e$ and the spectrum reaches its peak at $p \approx 1 a_e H_e$.  For $m_{3/2} > m_\phi \approx 2 H_e$ the high-momentum tail steepens to $p^{-15/2}$, which can be understood to result from kinematically blocking the 2-to-2 channel and giving way to the sub-dominant 4-to-2 channel~\cite{Ema:2018ucl,Ema:2019yrd,Chung:2018ayg,Basso:2022tpd}.  For larger values of the mass, see the curve labeled $m_{3/2} = 4 H_e \approx 2 m_\phi$, the spectrum softens to $p^2$ for small $p$, since particle production is not efficient enough to saturate the Pauli-blocking upper limit, and the amplitude decreases exponentially with larger mass. All of these behaviors have been observed previously for the CGPP of spin-$\onehalf$ particles (see, e.g., \cite{Kolb:2023ydq}).  

Now we'll discuss the helicity-$\onehalf$ polarization modes.  The results of our numerical calculation are presented in the right panel of \fref{fig:highmass_spectra}.  We observe many of the same feature that are present for the helicity-$\threehalf$ polarization modes, namely the low-momentum scaling with $p^3$ or $p^2$ and the high-momentum scaling with $p^{-3/2}$ or $p^{-15/2}$.  One new feature are the large-amplitude oscillations seen on the curves labeled $m_{3/2} = 1.0 H_e$ and $1.5 H_e$.  Oscillations such as these have been understood to arise from quantum interference effects~\cite{Basso:2022tpd}.  Another new feature is an intermediate rising behavior that scales as approximately $p^2$, and which is most pronounced on the curves labeled $m_{3/2} = 0.5 H_e$.  Consequently, reducing the raritron mass increases the amount of particle production, and in the next subsection we'll see how this behavior connects to the catastrophic production for low-mass raritrons.  

\begin{figure}[t]
    \centering
    \includegraphics[width=1.00\linewidth]{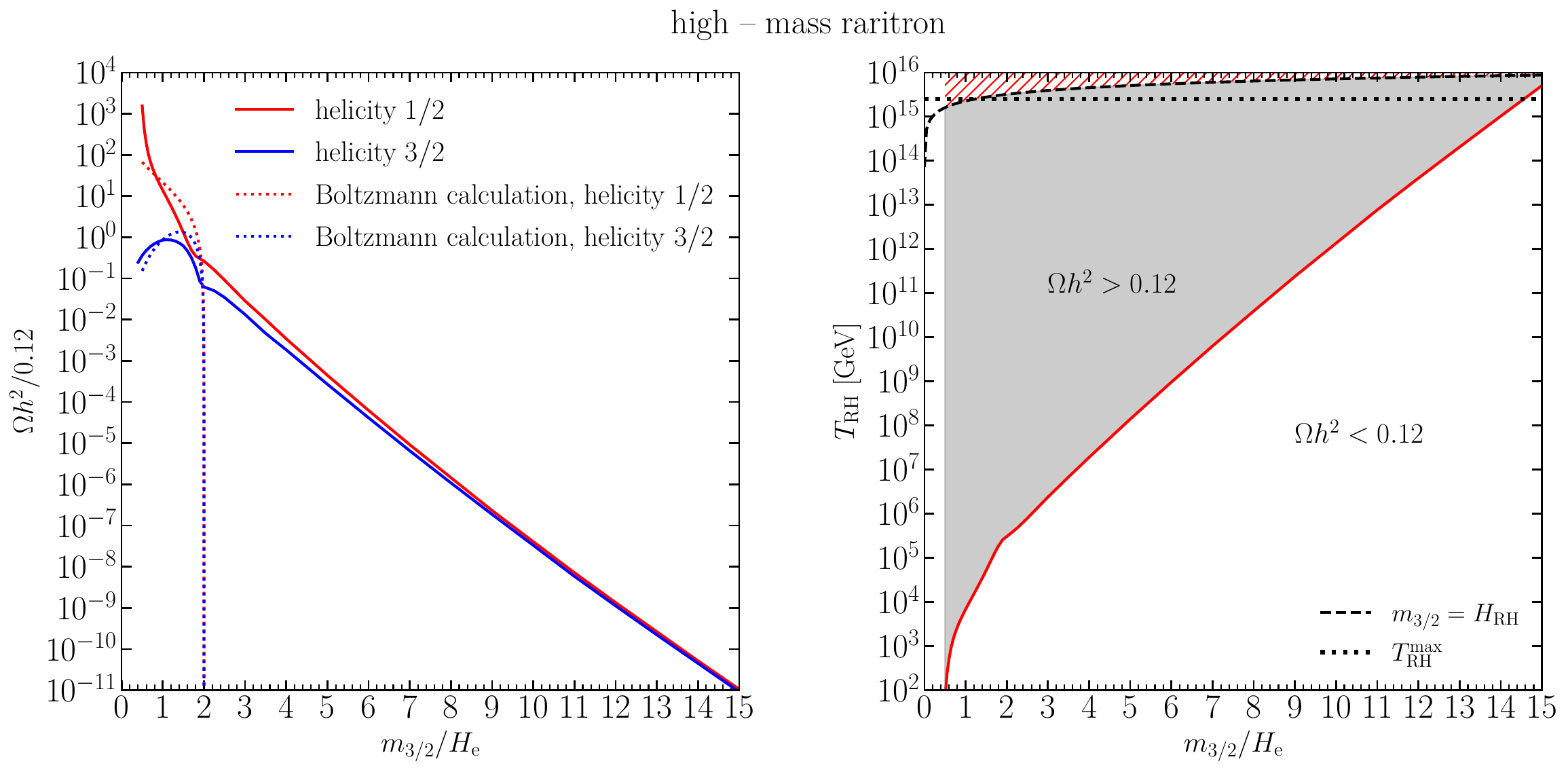}
    \caption{\label{fig:highmass_Oh2}
    Relic abundance and parameter space of high-mass raritron models.  \textit{Left:}  We plot the predicted relic abundances of helicity-1/2 and helicity-3/2 raritrons, $\Omega_{\onehalf} h^2$ and $\Omega_{\threehalf} h^2$, as a function of the raritron mass $m_{3/2}$ while assuming Quadratic Inflation for which the Hubble parameter at the end of inflation is $H_e \approx 8.5 \times 10^{12} \GeV$.  We take $T_\RH{} = 10^5 \GeV$ and more generally $\Omega h^2 \propto T_\RH{}$.  The dotted curves show the relic abundances predicted by the  2-to-2 Boltzmann calculation for $m_{3/2} < m_\phi \approx 2 H_e$.     \textit{Right:}  We vary the raritron mass $m_{3/2}$ and the reheating temperature $T_\RH{}$.  Assuming that raritrons are cosmologically long lived, we show the regions of parameter space where raritrons make up a subdominant fraction of the dark matter ($\Omega h^2 < 0.12$), regions where they saturate the dark matter abundance ($\Omega h^2 = 0.12$), and regions that are excluded by the overproduction of raritrons ($\Omega h^2 > 0.12$). We also indicate the maximum reheating temperature $T_\RH{}^\mathrm{max} \approx 2.5 \times 10^{15} \GeV$ consistent with energy conservation; see \sref{sub:inflation}.  In the red hatched region, the late-reheating assumption $H_\RH{} < m_{3/2}$ fails.
    }
\end{figure}

By way of summarizing the results for high-mass raritron models, we show the parameter space in \fref{fig:highmass_Oh2}.  Assuming that the raritron particles are cosmologically long-lived, we evaluate \eref{eq:Omegah2} to calculate their cosmological energy fraction today $\Omega$ as a function of the raritron mass $m_{3/2}$ and the reheating temperature $T_\RH{}$. Note that $\Omega h^2 = 0.120 \pm 0.001$ with $H_0 = 100 h \, \mathrm{km}\, \mathrm{sec}^{-1}\, \mathrm{Mpc}^{-1}$ is the dark matter abundance inferred by \textit{Planck}~\cite{Planck:2018vyg}.  On the left panel, we compare the prediction for $\Omega h^2$ derived from a numerical calculation of the spectrum using the Bogoliubov formalism against a prediction for $\Omega h^2$ based on the analytical Boltzmann formalism.  Although the Boltzmann formalism only captures the high-momentum tail of the spectrum, whereas $\Omega h^2$ is calculated by integrating over the full spectrum, for the high-mass raritron models with $0.4 m_\phi \lesssim m_{3/2} < m_\phi \approx 2 H_e$, Boltzmann provides a decent approximation for $\Omega h^2$. On the right panel, we show the regions of parameter space where raritrons make up a subdominant fraction of the dark matter ($\Omega h^2 < 0.12$), regions where they saturate the dark matter abundance ($\Omega h^2 = 0.12$), and regions that are excluded by the overproduction of raritrons ($\Omega h^2 > 0.12$).  We find that raritrons produced by CGPP could make up all of the dark matter for masses up to $m_{3/2} \approx 14.5 H_e$.  Heavier raritrons are necessarily underproduced, and lighter raritrons could be overproduced if the reheating temperature is too high. If the raritron is not stable, then one should investigate the phenomenological consequences of its decay.  Similar to the classic gravitino problem of supersymmetry~\cite{Krauss:1983ik,Ellis:1984eq}, the raritron's decay may cause problems for cosmological observables, such as the light element abundances and microwave background radiation.  

\subsection{Low-mass raritron}
\label{sub:low_mass}

Here we study the class of models with a low-mass raritron.  As we noted already at \eref{eq:raritron_models}, the low-mass raritron achieves $|c_s(\eta_n)| = 0$ at one or several times $\eta_n = \eta_1, \eta_2, \cdots$, and for Quadratic Inflation this condition corresponds to a small raritron mass, $m_{3/2} \lesssim 0.4 H_e$.  In the remainder of this subsection, first we'll discuss the spectrum of the helicity-$\threehalf$ polarization modes, then the helicity-$\onehalf$ polarization modes, and then the relic abundance constraints across parameter space.  

\begin{figure}[t]
    \centering
    \includegraphics[width=0.48\linewidth] {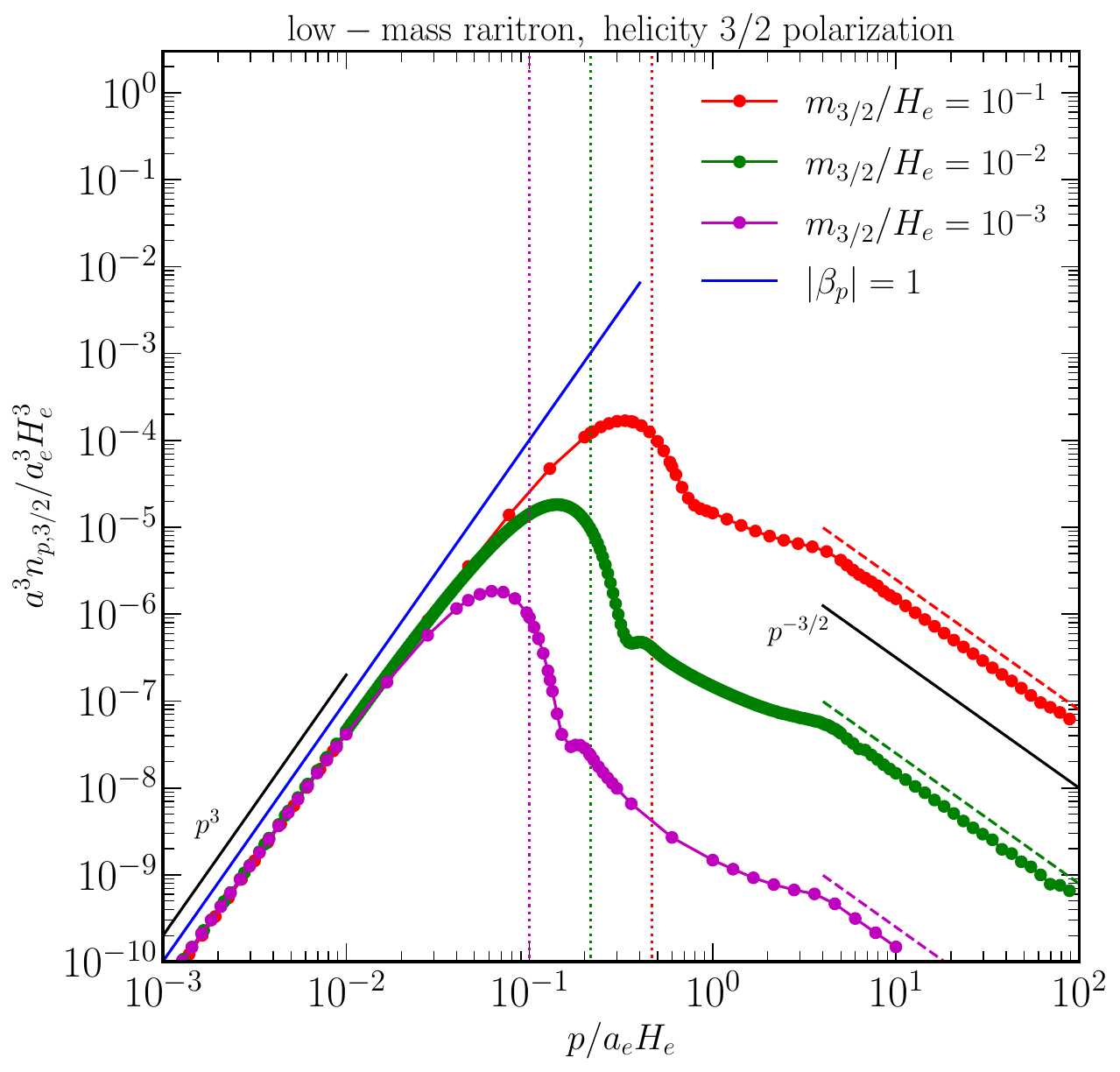}
    \hfill 
    \includegraphics[width=0.48\linewidth] {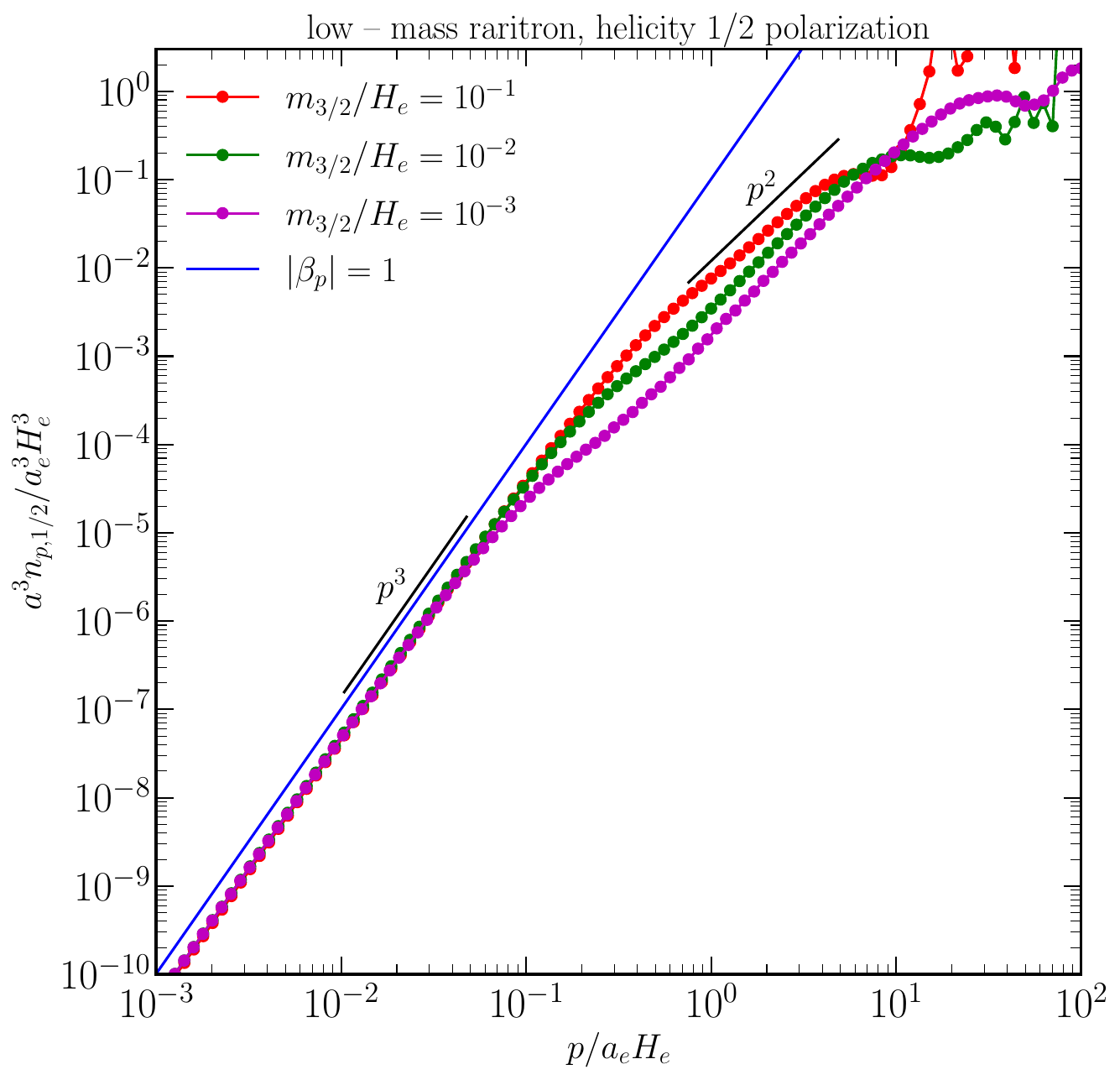}
    \caption{\label{fig:lowmass_spectra}
    Spectra of CGPP for low-mass raritron models.  Style and notation is the same as \fref{fig:highmass_spectra}.  
    }
\end{figure}

To begin we'll discuss the results for the helicity-$\threehalf$ polarization modes.  The results of our numerical calculation are presented in the left panel of \fref{fig:lowmass_spectra}.  The spectrum rises as $p^3$, reaches its maximum at approximately $p / a_e H_e = (m/H_e)^{1/3}$, decreases exponentially, and flattens into a $p^{-3/2}$ power law tail above $p \approx 4 a_e H_e$.  The spectrum in the low-momentum modes is $a^3 n_{p,\threehalf} \approx p^3 / 2\pi^2$, which corresponds to $|\beta_{p,\threehalf}|^2 \approx \tfrac{1}{2}$ in \eref{eq:a3np_from_beta2}.  Since Pauli blocking restricts $|\beta_{p,\threehalf}|^2 \leq 1$, a value of $1/2$ signals that the particle production is very efficient.  The same behavior has been observed previously for spin-$\onehalf$ particles~\cite{Chung:2011ck}, which obey the same equations of motion as helicity-$\threehalf$ raritrons. Toward large momentum, the spectrum breaks at $p \approx p_\ast = a_e H_e^{2/3} m_{3/2}^{1/3}$, which corresponds to the mode that enters the horizon at the same time as the field is released from Hubble drag~\cite{Chung:2011ck}; i.e., $p_\ast = a(\eta_\ast) m_{3/2}$ and $m_{3/2} = H(\eta_\ast) \approx H_e \, (a_\ast / a_e)^{-3/2}$.  The spectrum in the high-momentum modes goes as $p^{-3/2}$, and both the scaling and amplitude agree well with the analytical Boltzmann calculation for $p > a_* m_\phi \approx 4 a_e H_e$ where the Boltzmann calculation is valid.  However, it is important to emphasize that the full spectrum achieves its maximum $p \approx p_\ast = a_e H_e^{2/3} m_{3/2}^{1/3}$, where the analytical Boltzmann calculation is inapplicable.  If one were to only perform the Boltzmann calculation and set the spectrum to zero for $p < a_e m_\phi$, then one would underestimate the total amount of particle production.  The error would be a factor that grows as $(m_{3/2}/H_e)^{-1}$ and can be much larger than one for a light raritron.  

Next we'll discuss the results for the helicity-$\onehalf$ polarization modes.  
The results of our numerical calculation are presented in the right panel of \fref{fig:lowmass_spectra}.  The spectrum rises as $p^3$, softens into $p^2$ above $p \approx p_\ast = a_e H_e^{2/3} m_{3/2}^{1/3}$, and oscillates about a $p^2$ scaling above $p \approx 10 a_e H_e$.  The $p^3$ scaling is explained by Pauli blocking, as we have already discussed for the helicity-$\threehalf$ polarization modes.  The break at $p \approx p_\ast = a_e H_e^{2/3} m_{3/2}^{1/3}$ and the $p^2$ scaling is a consequence of the different redshift evolution for modes that are outside the horizon and inside the horizon.  We note that the softening to $p^2$ was overlooked in earlier studies of CGPP for low-mass raritrons~\cite{Hasegawa:2017hgd,Kolb:2021xfn,Kolb:2021nob,Hashiba:2022bzi}, which either explicitly claimed or implicitly assumed that the $n_p \propto p^3$ scaling would continue toward arbitrarily large momentum.  However, the break can also be seen in inspection of Fig.~1 of \rref{Kolb:2021xfn}.  From our numerical solution, it does appear that the $p^2$ scaling may continue indefinitely.  We do not observe the spectrum turning over for any $p$ up to $100 a_e H_e$, which is the largest momentum for which we can numerically solve the mode equations reliably.  If one assumes that the spectrum does not turn over toward larger $p$, then the total amount of particle production (calculated by integrating the spectrum over $\dd p / p$), is divergent.  This behavior has been noted previously \cite{Hasegawa:2017hgd} and called catastrophic particle production \cite{Kolb:2021xfn,Kolb:2021nob}.  However, it's worth clarifying that the earlier studies found $a^3 n_{p,\onehalf} \propto p^3$ corresponding to uniformly-efficient production of high-momentum modes $|\beta_{p,\onehalf}| \propto p^0$.  By contrast, we now find $a^3 n_{p,\onehalf} \propto p^2$, corresponding to less efficient production of high-momentum modes $|\beta_{p,\onehalf}| \propto p^{-1/2}$.  

\begin{figure}[t]
    \centering
    \includegraphics[width=1.00\linewidth]{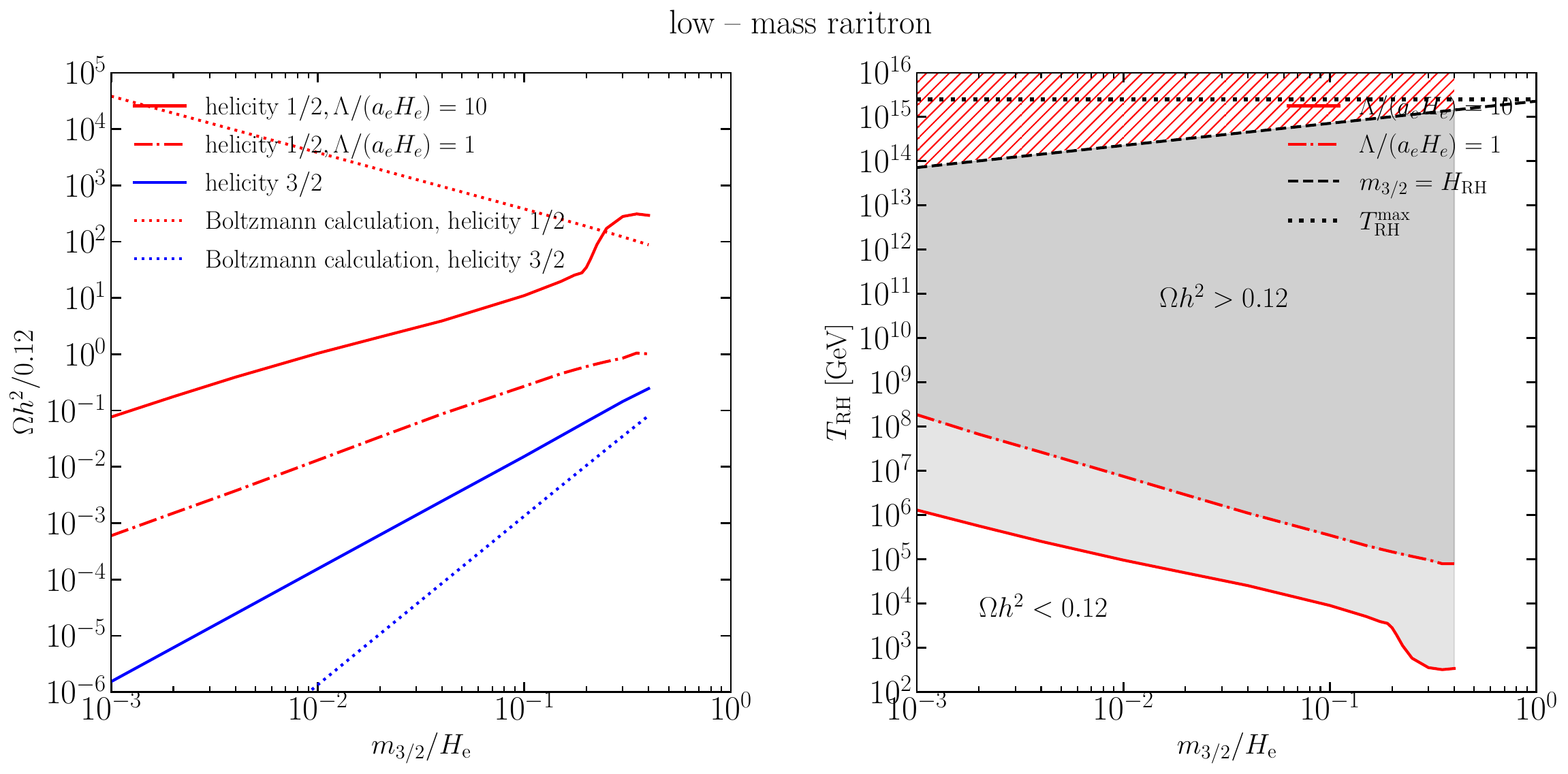}
    \caption{\label{fig:lowmass_Oh2}
    Relic abundance and parameter space of low-mass raritron models.  Style and notation is the same as \fref{fig:highmass_Oh2}.  For the helicity-$\onehalf$ polarization we illustrate two values for the UV cutoff on the comoving momentum $\Lambda = a_e H_e$ and $10 a_e H_e$.  
    }
\end{figure}

By way of summarizing the results for low-mass raritron models, we show the relic abundance and parameter space in \fref{fig:lowmass_Oh2}.  For the helicity-$\onehalf$ polarization modes, we introduce a hard cutoff on the comoving momentum by imposing $a^3 n_{\onehalf} = \int_0^\Lambda (\dd p/p) \, a^3 n_{p,\onehalf}$.  Lowering the momentum cutoff $\Lambda$ reduces the relic abundance by a factor $\Omega h^2 \propto \Lambda^2$, because we find that the spectrum scales as $p^2$; this can be contrasted with earlier work~\cite{Hasegawa:2017hgd,Kolb:2021xfn,Kolb:2021nob,Hashiba:2022bzi} that took the spectrum to be proportional to $p^3$ and found $\Omega h^2 \propto \Lambda^3$ instead.  For the helicity-$\threehalf$ polarization modes we find $\Omega h^2 \propto m_{3/2}^2$, since lowering the mass reduces the energy per particle by a factor of $m_{3/2}$ and it also reduces the efficiency of particle production by a factor proportional to $m_{3/2}$.  For the helicity-$\onehalf$ polarization modes we find $\Omega h^2 \propto m_{3/2}$, since the spectrum is insensitive to the mass.  Since the helicity-$\onehalf$ polarization modes experience catastrophic particle production, they are the larger contribution to $\Omega h^2$.  We present constraints on the parameter space for different assumptions about the UV cutoff $\Lambda$.  We show the regions of parameter space where low-mass raritrons are a subdominant component of the dark matter ($\Omega h^2 < 0.12$), regions where they can make up all of the dark matter ($\Omega h^2 = 0.12$), and regions that are excluded by the overproduction of low-mass raritrons ($\Omega h^2 > 0.12$). Note that low-mass raritrons can make up all the dark matter even if $m_{3/2}/H_e \ll 1$, provided that $T_\RH{}$ is sufficiently large.  Consider the choice $\Lambda/a_eH_e=10$ on the right-hand side of  \fref{fig:lowmass_Oh2}. For $m_{3/2}/H_e\lesssim 0.2$, the curve that results in $\Omega h^2=0.12$ is approximately $m_{3/2}/H_e =10^3(\mathrm{GeV}/T_\mathrm{RH})$. The intersection of the $\Omega h^2=0.12$ curve with the $H_{\rm RH}=m_{3/2}$ curve occurs at $m_{3/2} \simeq 5.75\times10^{-9} H_e \approx 5\times10^{4}\,\mathrm{GeV}$.  Light raritrons produced by CGPP could be the dark matter, and even lighter raritrons would be allowed to be the dark matter for a larger cutoff. 

\subsection{Evolving-mass raritron}
\label{sub:evolving_mass}

Here we study the class of models with an evolving-mass raritron.  As we noted already at \eref{eq:raritron_models}, the evolving-mass raritron maintains $|c_s(\eta)| = 1$ at all times, because the raritron mass $m_{3/2}(\eta)$ varies according to the ODE in \eref{eq:cs_condition}, which reads $( a^{-1} \partial_\eta m_{3/2} )^2 = ( 3 m_{3/2}^2 + H^2 - \tfrac{1}{6} R) ( 2 H^2 + \tfrac{1}{6} R )$.  However, this condition leaves the sign of $\partial_\eta m_{3/2}$ undetermined. The positive root has $\partial_\eta m_{3/2} > 0$ at all times, and it leads to a solution in which $m_{3/2}(\eta)$ diverges at late times. The negative root branch also leads to a divergent solution at late times.  In the context of supergravity, we expect the gravitino mass to be finite at all times, including during and after inflation, and is reasonable to expect the raritron to exhibit this property also.

\begin{figure}[h!]
    \centering
    \includegraphics[width=0.70\linewidth]{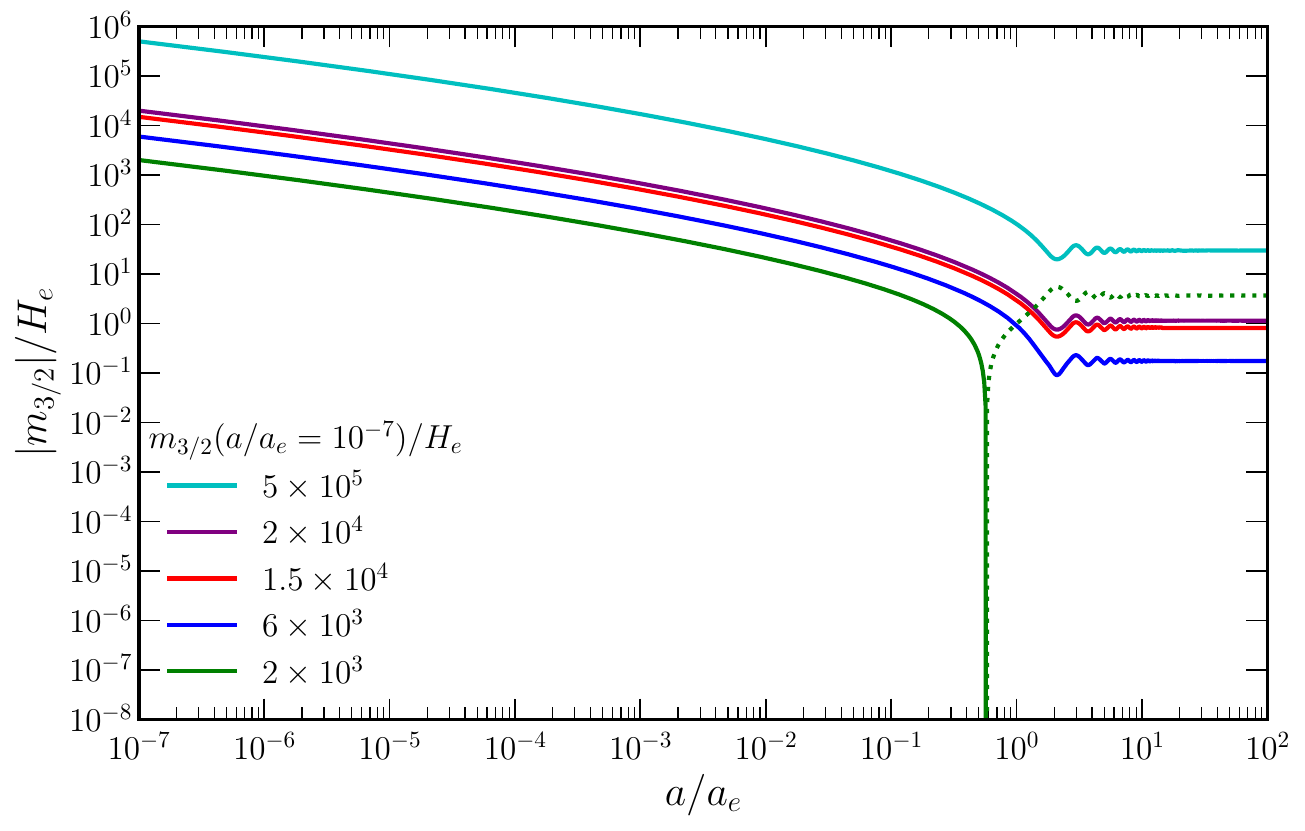}
    \caption{\label{fig:m32_vs_t}
    Evolution of the raritron mass $m_{3/2}$ with scale factor $a$.  The various curves correspond to different initial conditions for $m_{3/2}$ at $a/a_e = 10^{-7}$.  For the curve labeled $m_{3/2}/H_e = 2 \times 10^3$, the raritron mass passes through zero at $a/a_e \approx 0.6$ and afterward it is negative (indicated by the dotted curve).
    }
\end{figure}

In order to avoid a divergent raritron mass, and to remain agnostic regarding the UV completion of the raritron and of inflation (for example, into supergravity), we adopt a sign convention that ${\rm sign}(\partial_\eta{m}_{3/2} )= {\rm sign}(\dot{\phi})$ where $\phi$ is the inflaton of the quadratic inflation model. This prescription leads to a solution that settles to a constant at late times.  \fref{fig:m32_vs_t} shows the solution $m_{3/2}(\eta)$ for several initial conditions.  The raritron mass remains approximately constant during inflation, and for the parameters that we have plotted it is initially much larger than the inflationary Hubble scale $m_{3/2} \gg H_{60} > H_e$.  After inflation has ended ($a/a_e > 1$) and $\dot{\phi}$ is oscillating, the raritron mass also begins to oscillate around a constant.  At the end of reheating, the inflaton field has fully decayed, meaning $\phi(t) = 0$, and the universe enters radiation domination.  
Our analysis does not extend beyond reheating. 

\begin{figure}[h!]
    \centering
    \includegraphics[width=0.48\linewidth] {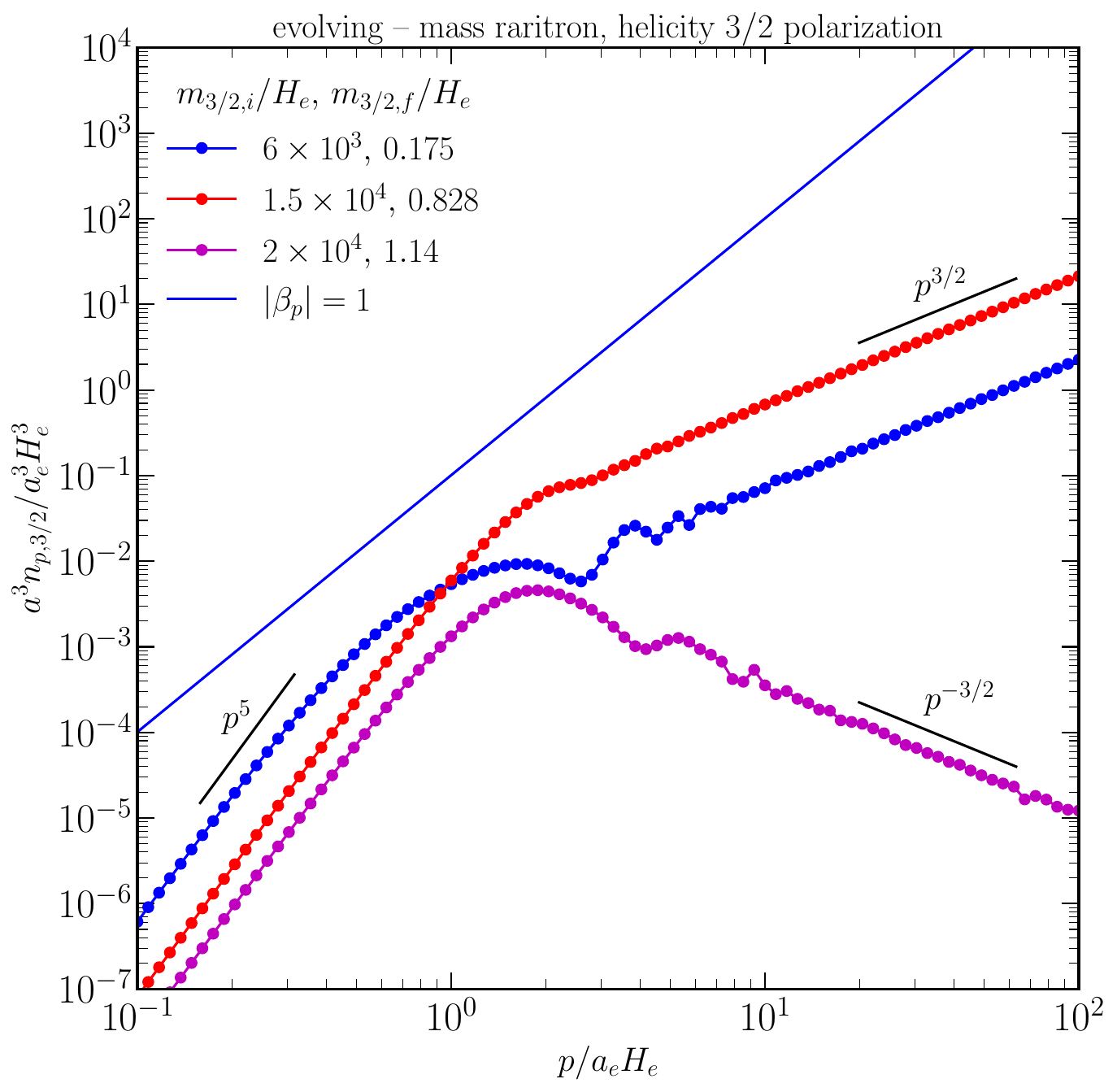}
    \hfill 
    \includegraphics[width=0.48\linewidth] {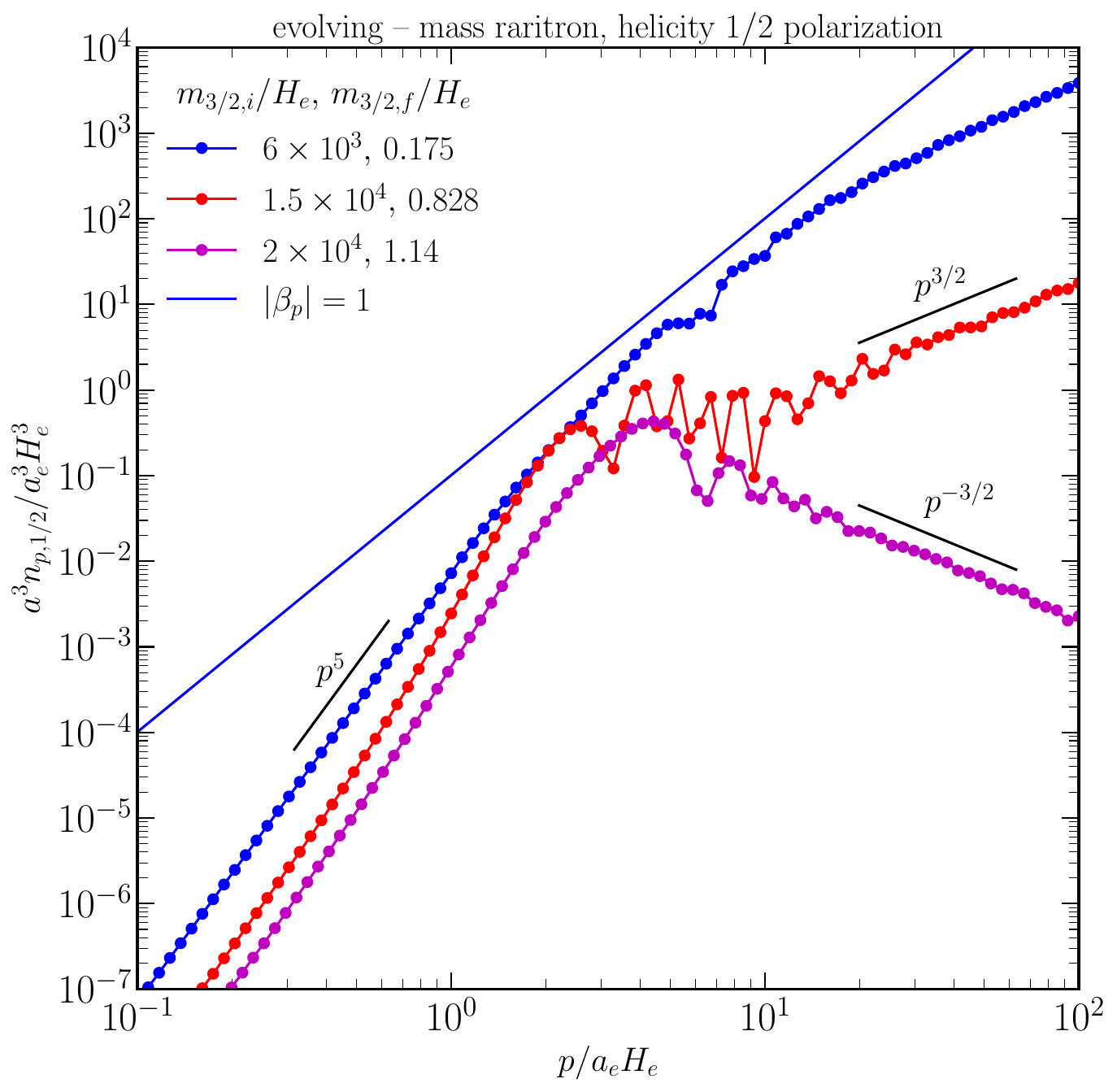}
    \caption{\label{fig:evolvemass_spectra}
    Spectra of CGPP for evolving-mass raritron models.  The various curves correspond to different values $m_{3/2,i} = m_{3/2}(a/a_e = 10^{-7})$ and final values $m_{3/2,f} = m_{3/2}(a/a_e = 10^3)$   \textit{Left:}  Helicity-$\threehalf$ polarization mode.  \textit{Right:}  Helicity-$\onehalf$ polarization mode. 
    }
\end{figure}

The results of our numerical calculation are presented in \fref{fig:evolvemass_spectra}. The spectra for both the helicity-$\threehalf$ and the helicity-$\onehalf$ polarization modes have the same scalings and breaks. The spectra rise as $p^5$, break at $p \sim a_e H_e$, and continue as either $p^{3/2}$ or $p^{-3/2}$. The two curves corresponding to $m_{3/2,f}/H_e < 1$ display a rising spectrum $\propto p^{3/2}$, and we do not observe the spectrum turning over for $p$ up to $100 \, a_e H_e$, which is the largest momentum for which we can numerically solve the mode equations reliably. The one curve corresponding to $m_{3/2,f}/H_e > 1$ displays a falling spectrum $\propto p^{-3/2}$.  

The universal $p^5$ behavior toward small momentum is a consequence of the changing raritron mass.  During inflation when modes of the raritron field are leaving the horizon, the raritron mass is much larger than the Hubble scale.  The suppresses the amplitude of the raritron field, and the suppression is increasingly strong for modes with smaller wavenumber $k$, which spend longer outside the horizon.  

The qualitatively different behavior for $m_{3/2,f}$ above or below $H_e$ can be traced to the post-inflationary particle production.  Recall that for Quadratic Inflation, the inflaton mass is $m_\phi \approx 2 H_e$.  So $m_{3/2,f} < H_e$ means $m_\phi \gtrsim 2 m_{3/2,f}$ and it is kinematically possible for the inflaton condensate to decay into pairs of raritrons after inflation.  One can also understand this behavior in the Bogoliubov formalism using \eref{eq:FT}.  Since $\mu_a(\eta) \propto m_{3/2}(\eta)$ and since $\partial_\eta m_{3/2}(\eta)$ flips sign each time $\dot{\phi}=0$ and since $\phi(t)$ oscillates at an angular frequency of $m_\phi$, it follows that the Fourier transform of $\mu_a^\ast/\omega_a$ picks out the component that oscillates at an angular frequency of $m_\phi$. For $m_{3/2,f} > H_e$ the leading channel is kinematically blocked, leaving behind the subleading scaling. 

More broadly speaking, it is illuminating to contrast the low-mass raritron and the evolving-mass raritron.  The low-mass raritron displays catastrophic particle production in the helicity-$\onehalf$ polarization modes, whose spectrum continues to rise as $a^3 n_{p,\onehalf} \propto p^2$.  In earlier work, this behavior was attributed to the low-mass raritron's vanishing sound speed after inflation, $|c_s(\eta_n)| = 0$.  This observation was partly our motivation to study the evolving-mass raritron, which has $|c_s(\eta)| = 1$ at all times.  We anticipated to find that the catastrophic production would be avoided, because of the constant sound speed.  However, we instead find that the catastrophic production persists for the evolving-mass raritron if its late-time mass is small $m_{3/2,f}/H_e < 1$.  The divergence is softened from $a^3 n_p \sim p^2$ to $p^{3/2}$, and it is present in both the helicity-$\threehalf$ and the helicity-$\onehalf$ polarization modes.  Since the helicity-$\threehalf$ mode equation is independent of $c_s(\eta)$, the newfound singular behavior can only be traced to the time-dependent raritron mass $m_{3/2}(\eta)$.  Since $m_{3/2}(\eta)$ is varied ``by hand'' it is perhaps not too surprising that one can find enhanced particle production.  It's also worth noting that $a^3 n_p \propto p^{3/2}$ corresponds to $|\beta_k| \propto k^{-3/4}$, which is smaller at larger $k$.  A mode with wavevector $\kvec$ is less efficiently produced if $k = |\kvec|$ is larger, but since there are more modes at higher $k$, the spectrum $a^3 n_p \propto p^3 |\beta_k|^2$ rises.  

For some values of the raritron mass, the comoving number density spectrum is rising as $p^{3/2}$ toward large comoving momentum.  
Of course this scaling cannot continue toward arbitrarily large $p$, because it would correspond to infinite particle production.  
There are several physical effects that enter to truncate the rising spectrum and to impose a UV cutoff on the comoving momentum.  

An interacting massive spin-3/2 particle is strongly coupled in the UV above an energy scale of $E \sim \sqrt{m_{3/2} M_\ast}$ where $m_{3/2}$ denotes the mass and $M_\ast^{-1}$ denotes the  interaction strength.  
For example, the authors of \rref{Melville:2019tdc} studied a model with dimension-6 quartic self-interactions $\mathcal{L}_\mathrm{int} \sim m_{3/2} \psi^4 / M_\ast^3$.  
Since our perturbative calculation break downs at energies for which the raritron is strongly coupled, we should disregard the spectrum at $p > \Lambda$.  
If we take $M_\ast = M_\mathrm{Pl}$ the strong coupling limit at the end of inflation implies $\Lambda = a_e \sqrt{m_{3/2} M_\mathrm{Pl}}$.  
Note that $\Lambda / a_e H_e = \sqrt{m_{3/2} M_\mathrm{Pl} / H_e^2} \approx 500 (m_{3/2}/H_e)^{1/2} (H_e / 8.5 \times 10^{12} \; \mathrm{GeV})^{-1/2}$.  
On \fref{fig:evolvemass_spectra} we truncate the plot at $p / a_e H_e = 100$, so the range of comoving momenta shown there is below the strong coupling scale. 

When we calculate the spectrum of particles that arise through CGPP, we take the inflationary background to be fixed.  
In other words, we neglect the back reaction of the particle production on the inflaton field's evolution.  
This approach is an excellent approximation as long as the energy carried by the gravitationally produced particles is much smaller than the energy carried by the inflaton field.  
However, if the rising spectra in \fref{fig:evolvemass_spectra} were to continue towards arbitrarily large $p$ then the rariton's energy would exceed the inflaton's energy and invalidate our approach of neglecting the backreaction.  Conversely, to estimate when the backreaction \textit{can} be neglected, suppose that the rising spectrum is cutoff at a comoving momentum $p = \Lambda$ defined by $3 M_\mathrm{pl}^2 H_e^2 = m_{3/2} n_{3/2}(t_e)$, which resolves to approximately $\Lambda / a_e H_e \approx 10^5 (m_{3/2} / H_e)^{-2/3} (H_e / 8.5 \times 10^{12} \; \mathrm{GeV})^{-2/3}$.  
For the parameters shown on \fref{fig:evolvemass_spectra}, the spectrum's cutoff may fall outside the range of $p$ shown. 

\section{Conclusion}
\label{sec:conc}

In this work we have studied the creation of massive spin-$\threehalf$ particles during and at the end of inflation due to their gravitational interactions alone.  The phenomenon of cosmological gravitational particle production provides a simple and unavoidable origin for these raritrons in the early universe.  Assuming that they are stable, or at least cosmologically long lived, we calculate their cosmological energy fraction today and show that they can make up all or some of the dark matter across a wide range of parameter space.  Conversely, models in which raritrons are overproduced suffer from a cousin of the Cosmological Gravitino Problem.  

We propose three different model classes that we call high-mass raritron, low-mass raritron, and evolving-mass raritron.  Their definitions can be found in \eref{eq:raritron_models}.  For each model we calculate the comoving momentum spectra, $a^3 n_{p,\threehalf}$ and $a^3 n_{p,\onehalf}$, for both the helicity-$\threehalf$ and the helicity-$\onehalf$ polarization modes.  We perform these calculations using both the Bogoliubov formalism and (when applicable) the Boltzmann formalism.  Several of our key findings can be summarized as follows:  
\begin{itemize}
    \item  For the high-mass raritron, the Bogoliubov-derived spectrum is peaked at $p \approx a_e H_e$, which is on the edge of validity of Boltzmann calculation.  As one can see from \fref{fig:highmass_Oh2}, the Boltzmann calculation only underestimates the Bogoliubov calculation by an order one factor.  
    \item  For the low-mass raritron's helicity-$\threehalf$ polarization modes, the Bogoliubov-derived spectrum is peaked at $p \approx a_e H_e^{2/3} m_{3/2}^{1/3}$, which can be far below the validity of Boltzmann calculation if $m_{3/2} \ll H_e$.  As one can see from \fref{fig:lowmass_Oh2}, the Boltzmann calculation underestimates the Bogoliubov calculation by a factor of order $m_{3/2}/H_e$, which can be very far from order one.  
    \item  For the low-mass raritron's helicity-$\onehalf$ polarization modes, we find that the spectrum rises as $a^3 n_{p,\onehalf} \propto p^2$ toward large comoving momentum.  If the rising spectrum would continue to arbitrarily large $p$, then the total comoving number density would be divergent; this behavior has been called catastrophic particle production.  Regulating the integral with a hard cutoff $p \leq \Lambda$ leads to $a^3 n_{\onehalf} \propto \Lambda^2$.  Earlier work had claimed $a^3 n_{p,\onehalf} \propto p^3$ and $a^3 n_{\onehalf} \propto \Lambda^3$ instead. Avoiding the overproduction of low-mass raritrons implies a strong upper limit on the reheating temperature; see \fref{fig:lowmass_Oh2}.  
    \item  We propose the evolving-mass raritron class of models, which are defined by a time-dependent mass parameter $m_{3/2}(\eta)$ such the effective sound speed maintains $|c_s(\eta)|=1$ at all times.  In this way, the gradient instability in the low-mass raritron's helicity-$\onehalf$ polarization mode is avoided.  
    \item  For the evolving-mass raritron, we find that the spectrum is falling toward high momentum for a model in which the mass approaches a constant $m_{3/2} \approx 1.14 H_e$ at late times.  However, for a pair of models in which $m_{3/2} < H_e$ at late times, we again encounter a rising spectrum $a^3 n_p \propto p^{3/2}$.  This study indicates that the rising spectrum need not arise from vanishing sound speed $c_s(\eta) = 0$, but it can also result from an oscillating mass $m_{3/2}(\eta)$.  
\end{itemize}
Note that all of our numerical results assume Quadratic Inflation with $m_\phi = 1.7 \times 10^{13} \GeV$.  Other models of inflation with comparable values for the inflationary Hubble scale would lead to a similar level of particle production.  
A smaller Hubble scale would imply less efficient particle production, and our conclusions should be adapted accordingly.  

In this work we have focused on the implications of CGPP for raritron dark matter.  
In other contexts, however, the particles produced by CGPP may give rise to possibly detectable secondary gravitational waves~\cite{Ananda:2006af,Baumann:2007zm,Figueroa:2017vfa,Ebadi:2023xhq,Chakraborty:2024rgl,Kumar:2024hsi,Chakraborty:2025oyj,Garcia:2025yit,Garcia:2025wmu}.  
It would be interesting to calculate the gravitational wave spectrum that is expected to arise from raritron CGPP, particularly for the low-mass and evolving-mass raritron models that display a UV-enhanced spectrum.  
Since measurements of the CMB place an upper bound on the cosmological abundance of dark radiation (\ie{}, $\Delta N_\mathrm{eff}$), an overly-efficient production of gravitational waves would imply additional constraints on raritron dark matter produced via CGPP.  

In this work we have modeled the raritron as the quantum excitation of a spin-$\threehalf$ Rarita-Schwinger field that is minimally coupled to gravity.  The most notable context in which a massive spin-$\threehalf$ particle arises is arguably supergravity, where the spin-$\threehalf$ gravitino is superpartner to the spin-2 graviton.  Some models of supergravity map directly to the raritron model, whereas other simple models of supergravity entail additional dynamics and interactions that live outside the raritron model.  For example, the gravitino may mix with the inflatino, the fermionic superpartner to the inflaton, and these dynamics lead to a kinetic mixing matrix with unit eigenvalues.  In this work, we proposed the evolving-mass raritron, which has  $|c_s(\eta)|=1$ at all times, to assess what role the varying sound speed plays in CGPP.  Even though the sound speed is fixed, we nevertheless find a rising spectrum of raritrons, due instead to the oscillating time-dependent mass.  We view this calculation as a stepping stone toward a calculation of CGPP in models of supergravity, which we hope to pursue in future work.  

\acknowledgments
The authors thank Sarunas Verner for helpful discussions of the Boltzmann calculation, and we thank Takahiro Terada for comments on the draft.  This material is based upon work supported (in part: A.J.L.) by the National Science Foundation under Grant No.~PHY-2412797.  E.M. is supported in part by a Discovery Grant from the Natural Sciences and Engineering Research Council of Canada, and by a New Investigator Operating Grant from Research Manitoba.  We are grateful to the Munich Institute for Astro-, Particle and BioPhysics (MIAPbP), which is funded by the Deutsche Forschungsgemeinschaft (DFG, German Research Foundation) under Germany´s Excellence Strategy – EXC-2094 – 390783311, for hosting a workshop at which this work began. 

\appendix 
\section{Boltzmann Calculation}
\label{app:Boltzmann}

In this appendix we present a derivation of the spectrum of gravitationally-produced spin-$\threehalf$ particles using the Boltzmann formalism.  We consider 2-to-2 scattering mediated by an $s$-channel graviton exchange.  We adopt the matrix elements that were derived in an earlier study \cite{Kaneta:2023uwi}, which also derived expressions for the raritron relic abundance.  We extend the earlier work by also providing expressions for the raritron spectrum and by generalizing the sound speed to $c_s \neq 1$. 

\subsection{Sound speed unity}

Consider the reaction 
\begin{align}
    \phi(\kvec_{\phi_1}) + \phi(\kvec_{\phi_2}) \to \chi(\kvec_{\chi_1}, s_{\chi_1}) + \chi(\kvec_{\chi_2}, s_{\chi_2}) 
    \;,
\end{align}
where an inflaton $\phi$ with 3-momentum $\kvec_{\phi_1}$ is incident on an inflaton $\phi$ with 3-momentum $\kvec_{\phi_2}$, and they annihilate producing a pair of recoiling raritrons, the first with 3-momentum $\kvec_{\chi_1}$ and helicity $s_{\chi_1}$ and the second with 3-momentum $\kvec_{\chi_2}$ and helicity $s_{\chi_2}$.  This reaction contributes to the collision term in the raritron kinetic equation as 
\begin{align}\label{eq:dndt}
    \dot{n}_\chi + 3H n_\chi & = C 
    \;,
\end{align}
where $H(t) = \dot{a}/a$ is the cosmological expansion rate (\ie{}, Hubble rate), and where $n_\chi(t)$ is the number density of raritrons at time $t$, regardless of their helicity.  The collision integral is \cite{Gondolo:1990dk} 
\begin{align}\label{eq:Rphi2}
    C(t) & = 
    2 \, 
    \biggl( \frac{1}{2} 
    \int \! \frac{\dd^3\kvec_{\phi_1}}{(2\pi)^3} \frac{1}{2E_{\phi_1}} 
    \int \! \frac{\dd^3\kvec_{\phi_2}}{(2\pi)^3} \frac{1}{2E_{\phi_2}} \biggr) 
    \\ & \quad \times
    \biggl( \frac{1}{2} 
    \int \! \frac{\dd^3\kvec_{\chi_1}}{(2\pi)^3} \frac{1}{2E_{\chi_1}} \sum_{s_{\chi_1} = \pm 3/2, \pm 1/2}
    \int \! \frac{\dd^3\kvec_{\chi_2}}{(2\pi)^3} \frac{1}{2E_{\chi_2}} \sum_{s_{\chi_2} = \pm 3/2, \pm 1/2} \biggr)
    \nn & \quad \times
    f_\phi(\kvec_{\phi_1}) \, 
    f_\phi(\kvec_{\phi_2}) \, 
    \bigl| {\cal{M}}_{\phi\phi\to\chi\chi}(\kvec_{\phi_1}; \, \kvec_{\phi_2}; \, \kvec_{\chi_1}, s_{\chi_1}; \, \kvec_{\chi_2}, s_{\chi_2}) \bigr|^2 \, 
    \nn & \quad \times 
    (2\pi)^4 \delta^4(k_{\phi_1}+k_{\phi_2}-k_{\chi_1}-k_{\chi_2}) 
    \;.
    \nonumber
\end{align}
The inverse process $\chi\chi \to \phi\phi$ can be neglected since the $\chi$ abundance remains small.  The factor of $2$ accounts for the production of $\chi$ particles in pairs.  The factors of $1/2$ correct for the double counting that arises from integrating over the momenta of identical particles. The inflaton's one-particle phase space distribution function is denoted by $f_\phi(\kvec)$.  The Lorentz-invariant matrix element is denoted by $\mathcal{M}_{\phi\phi\to\chi\chi}$.  Since the helicity-$\threehalf$ and the helicity-$\onehalf$ raritrons each separately couple pairwise to gravity, it follows that the matrix element vanishes unless $|s_{\chi_1}| = |s_{\chi_2}|$, which means that the densities of helicity-$\onehalf$ and helicity-$\threehalf$ particles evolve independently.  This observation lets us write 
\begin{align}\label{eq:dndt_a}
    \dot{n}_{\chi,a} + 3H n_{\chi,a} & = C_{a}
    \;,
\end{align}
for $a = 3/2$ or $1/2$.  Here $n_{\chi,3/2}(t)$ is the number density of raritrons with either $s_\chi = 3/2$ or $-3/2$, and $n_{\chi,1/2}(t)$ is the density with $s_\chi = 1/2$ or $s_\chi = -1/2$.  The collision integrals, $C_{3/2}(t)$ and  $C_{1/2}(t)$, are calculated similarly to $R^{\phi^2}(t)$ in \eref{eq:Rphi2}, except that only the appropriate helicities are summed.  

The helicity-summed matrix element was calculated by the authors of \rref{Kaneta:2023uwi}, and we adopt their results.  Of particular importance is the kinematic point denoted by $\star$
\begin{align}
    \star: \quad 
    \kvec_{\phi_1} = \kvec_{\phi_2} = {\bm 0} 
    \ , \quad 
    \kvec_{\chi_2} = - \kvec_{\chi_1} 
    \ , \quad 
    E_{\chi_1} = E_{\chi_2} = m_\phi 
    \ , \quad 
    s = 4 m_\phi^2 
    \ , \quad 
    t = u = m_\chi^2 - m_\phi^2 
    \;,
\end{align}
which we denote by the symbol $\star$.  They define the helicity-summed squared matrix element as
\begin{align}
    \bigl| \overline{\mathcal{M}} \bigr|^2 = \frac{1}{4} \sum_{s_{\chi_1} = \pm 3/2, \pm 1/2} 
    \sum_{s_{\chi_2} = \pm 3/2, \pm 1/2} 
    \bigl| {\cal{M}}_{\phi\phi\to\chi\chi}(\kvec_{\phi_1}; \, \kvec_{\phi_2}; \, \kvec_{\chi_1}, s_{\chi_1}; \, \kvec_{\chi_2}, s_{\chi_2}) \bigr|^2 \biggr|_\star
    \;,
\end{align}
where the helicities are summed and the momenta are evaluated at $\star$.  
Similarly, we define 
\begin{align}\label{eq:Mbarsq_32_12}
    \bigl| \overline{\mathcal{M}}_{3/2,a} \bigr|^2 & = \frac{1}{4} \sum_{s_{\chi_1} = \pm a} 
    \sum_{s_{\chi_2} = \pm a} 
    \bigl| {\cal{M}}_{\phi\phi\to\chi\chi}(\kvec_{\phi_1}; \, \kvec_{\phi_2}; \, \kvec_{\chi_1}, s_{\chi_1}; \, \kvec_{\chi_2}, s_{\chi_2}) \bigr|^2 \biggr|_\star
    \;.
\end{align}
for $a = 3/2$ or $1/2$.  An expression for $|\overline{\mathcal{M}}|^2$ appears in Eq.~(19) of \rref{Kaneta:2023uwi}.  The corresponding matrix elements for each helicity can be inferred from their Eqs.~(34)~and~(35) to be 
\begin{subequations}\label{eq:msquared}
\begin{align}
    |\overline{\mathcal{M}}_{3/2,3/2}|^2 & 
    = \dfrac{1}{8} \dfrac{m_\phi^4}{ \MPl^4} \bigl( 1-r^2 \bigr) \, r^2 
    \;, \\ 
    |\overline{\mathcal{M}}_{3/2,1/2}|^2 & 
    = \dfrac{2}{9} \dfrac{m_\phi^4}{\MPl^4} \bigl( 1-r^2 \bigr) r^{-2} 
    \biggl( 1 - \dfrac{3}{4} r^2 \biggr)^2 
    \;,
\end{align}
\end{subequations}
where $r=m_{3/2}/m_\phi$. Note that $|\overline{\mathcal{M}}_{3/2,3/2}|^2 + |\overline{\mathcal{M}}_{3/2,1/2}|^2 = |\overline{\mathcal{M}}|^2$, \ie{} it sums to Eq.~(19) of \rref{Kaneta:2023uwi}.  

To evaluate the collision integral we approximate the inflaton's phase space distribution function as $f_\phi(\kvec) \approx n_\phi (2\pi)^3 \delta(\kvec)$, where $n_\phi(t)$ is the number density of inflaton particles at time $t$.  Consequently $\kvec_{\phi_1} = \kvec_{\phi_2} = {\bm 0}$, and both incident inflaton particles have the same momentum.\footnote{In the subset of the integration domain where $\kvec_{\phi_1} = \kvec_{\phi_2}$, the factor of $1/2$ that was added to correct for the double counting is not needed.  However, the domain of integration where $\kvec_{\phi_1} = \kvec_{\phi_2}$ is a set of measure zero.  So if the integrand were not singular here, its contribution to the integral would be zero, and no error would arise by including the unnecessary factor of $1/2$.  Although we approximate $f_\phi(\kvec) \propto \delta(\kvec)$ to evaluate the integral simply, we have in mind a non-singular function that is sharply peaked around $\kvec = {\bm 0}$.  Consequently, the factor of $1/2$ is still needed to correct for the double counting.  } Separating out the $|s_\chi| = 3/2$ and $1/2$ terms of the helicity sums and using \eref{eq:Mbarsq_32_12} leads to 
\begin{align}\label{eq:Rphisq_a_steps}
    C_a(t) 
    & = 
    2 \, 
    \biggl( \frac{1}{2} 
    \int \! \frac{\dd^3\kvec_{\phi_1}}{(2\pi)^3} \frac{1}{2E_{\phi_1}} 
    \int \! \frac{\dd^3\kvec_{\phi_2}}{(2\pi)^3} \frac{1}{2E_{\phi_2}} \biggr) 
    \biggl( \frac{1}{2} 
    \int \! \frac{\dd^3\kvec_{\chi_1}}{(2\pi)^3} \frac{1}{2E_{\chi_1}} 
    \int \! \frac{\dd^3\kvec_{\chi_2}}{(2\pi)^3} \frac{1}{2E_{\chi_2}} \biggr)
    \\ & \quad \times
    n_\phi (2\pi)^3 \delta(\kvec_{\phi_1}) \, 
    n_\phi (2\pi)^3 \delta(\kvec_{\phi_2}) \, 
    4 \bigl| \overline{\cal{M}}_{3/2,a} \bigr|^2 \, 
    (2\pi)^4 \delta^4(k_{\phi_1}+k_{\phi_2}-k_{\chi_1}-k_{\chi_2}) \nn 
    & = 
    2 \biggl( \frac{n_\phi}{2m_\phi} \biggr)^2 
    \int \! \frac{\dd^3\kvec_{\chi_1}}{(2\pi)^3} 
    \biggl( \frac{1}{2E_{\chi_1}} \biggr)^2 
    \bigl| \overline{\cal{M}}_{3/2,a} \bigr|^2 \, 
    (2\pi) \delta(2 m_\phi - 2 E_{\chi_1}) 
    \Bigr|_{\kvec_{\phi_1} = \kvec_{\phi_2} = {\bm 0}, \, \kvec_{\chi_2} = - \kvec_{\chi_1}} \nn 
    & = 
    2 \frac{(m_\phi n_\phi)^2}{m_\phi^2} \, 
    \frac{\bigl| \overline{\cal{M}}_{3/2,a} \bigr|^2}{32 \pi m_\phi^2} \int_0^\infty \! \! \dd k_{\chi_1} \,  
    \frac{k_{\chi_1}^2}{E_{\chi_1}^2} \, 
    \delta(m_\phi - E_{\chi_1}) 
    \Bigr|_{\kvec_{\phi_1} = \kvec_{\phi_2} = {\bm 0}, \, \kvec_{\chi_2} = - \kvec_{\chi_1}} \nn 
    & = 
    2 \biggl( \frac{n_\phi}{2m_\phi} \biggr)^2 
    \biggl( \frac{4\pi}{(2\pi)^3} \biggr) 
    \int_{m_\phi}^\infty \! \! \dd E_{\chi_1} \, \frac{k_{\chi_1} E_{\chi_1}}{4 E_{\chi_1}^2} 
    \bigl| \overline{\cal{M}}_{3/2,a} \bigr|^2 \, 
    (2\pi) \delta(2 m_\phi - 2 E_{\chi_1}) 
    \Bigr|_{\kvec_{\phi_1} = \kvec_{\phi_2} = {\bm 0}, \, \kvec_{\chi_2} = - \kvec_{\chi_1}} \nn 
    & = 
    2 \biggl( \frac{n_\phi}{2m_\phi} \biggr)^2 
    \biggl( \frac{4\pi}{(2\pi)^3} \biggr) 
    \frac{k_{\chi_1} m_\phi}{4 m_\phi^2} 
    \bigl| \overline{\cal{M}}_{3/2,a} \bigr|^2 \, 
    (2\pi) \frac{1}{2} 
    \Theta(m_\phi - m_\chi) 
    \Bigr|_{\star} \nn 
    & = 
    2 n_\phi^2(t) 
    \frac{\bigl| \overline{\cal{M}}_{3/2,a} \bigr|^2}{32 \pi m_\phi^2} 
    (1-r^2)^{1/2} \, 
    \Theta(m_\phi - m_\chi) 
    \;, 
    \nonumber
\end{align}
where $a = 3/2$ or $1/2$.  This expression is a factor of $2$ larger than Eq.~(20) of \rref{Kaneta:2023uwi}.  Their footnote~5 explains that the predicted rate differs depending on how one models the inflaton field in the initial state.  Treating the initial state as a collection of nonrelativistic inflaton particles (as we have done here) leads to a rate that is larger by a factor of $2$ as compared with treating the initial state as a coherent inflaton condensate.  Since we view the inflaton condensate as a more faithful model of the initial state, we remove the factor of $2$ and take the rate to be 
\begin{align}\label{eq:Rphisq_a}
    C_a(t) & = 
    n_\phi^2(t) 
    \frac{\bigl| \overline{\cal{M}}_{3/2,a} \bigr|^2}{32 \pi m_\phi^2} 
    (1-r^2)^{1/2} \, 
    \Theta(m_\phi - m_\chi) 
    \;,
\end{align}
for $a = 3/2$ or $1/2$.  

The comoving number density of gravitationally produced raritrons is calculated by integrating the kinetic equations \pref{eq:dndt_a}, 
\begin{align}
    a^3 n_{\chi,a} 
    & = \int_{t_\ast}^{t_\RH{}} \! \dd t^\prime \, a^3(t^\prime) \, C_{a}(t^\prime) 
    \;,
\end{align}
from the start of inflaton oscillations at time $t = t_\ast$ to the end of reheating at time $t = t_\RH{}$.  In general $t_e \lesssim t_\ast < t_\RH{}$, and we take $a_\ast = 2 a_e$ when numerical values are needed. 
The collision integral carries a time dependence through the factor $n_\phi(t)$.  During the epoch of reheating, but before an appreciable fraction of inflaton particles have decayed, we can approximate $n_\phi(t) \approx n_{\phi,e} [a(t) / a_e]^{-3}$ where $n_{\phi,e} \approx \rho_{\phi,e} / m_\phi \approx 3 \Mpl^2 H_e^2 / m_\phi$ using the Friedmann equation.  
Evaluating the integral gives 
\begin{align}
    \frac{a^3 n_{\chi,a}}{a_e^3 H_e^3} 
    & 
    = 
    \frac{6 \Mpl^4 H_e^4}{m_\phi^2} 
    \frac{\bigl| \overline{\cal{M}}_{3/2,a} \bigr|^2}{32 \pi m_\phi^2} 
    (1-r^2)^{1/2} \, 
    \Theta(m_\phi - m_\chi) 
    \frac{1}{H_e^4} \, 
    \biggl( 1 - \frac{a_\ast^{3/2}}{a_\RH{}^{3/2}} \biggr) \frac{a_e^{3/2}}{a_\ast^{3/2}} 
    \;. 
\end{align}
Upon summing $a=3/2$ and $1/2$ and setting $a_\ast = a_e$, the result is equivalent to Eq.~(43) of \rref{Kaneta:2023uwi}.  

A similar calculation gives the momentum spectrum of gravitationally produced raritrons.  We denote the spectrum as $a^3 n_{\chi,p,a} = p \dd (a^3 n_{\chi,a}) / \dd p$ where $\pvec = a \kvec$ is the comoving momentum of a particle with physical momentum $\kvec$, and where $p = |\pvec|$.  The spectrum satisfies a kinetic equation that is similar to \eref{eq:dndt_a},  
\begin{align}
    \frac{\dd}{\dd t} \bigl( a^3 n_{\chi,p,a} \bigr) 
    & = 
    n_\phi^2(t) \, 
    \frac{\bigl| \overline{\cal{M}}_{3/2,a} \bigr|^2}{32 \pi m_\phi^2} 
    \frac{p^3 a}{p^2 / a^2 + m_\chi^2} \, 
    \delta\Bigl( m_\phi - \sqrt{p^2/a^2 + m_\chi^2} \Bigr)
    \;,
\end{align}
where we have used an intermediate form of \eref{eq:Rphisq_a} and set $k_{\chi_1} = p/a$.  Integrating from time $t = t_\ast$ to time $t = t_\RH{}$ gives 
\begin{align}
    \frac{a^3 n_{\chi,p,a}}{a_e^3 H_e^3} & = 
    \frac{9 \Mpl^4 H_e^4}{m_\phi^2} \, 
    \frac{\bigl| \overline{\cal{M}}_{3/2,a} \bigr|^2}{32 \pi m_\phi^2} \, 
    (1-r^2)^{1/2} \, 
    \Theta(m_\phi - m_\chi) 
    \\ & \quad \times 
    \frac{(1-r^2)^{3/4}}{H_e^4} 
    \biggl( \frac{p}{a_e m_\phi} \biggr)^{\!\!-3/2} 
    \Theta\bigl( \sqrt{1-r^2} \, a_\ast m_\phi < p < \sqrt{1-r^2} \, a_\RH{} m_\phi \bigr) 
    \;, 
    \nonumber 
\end{align}
which scales $\propto p^{-3/2}$.  
Using the expressions for the squared matrix elements from \eref{eq:msquared} gives 
\begin{subequations}\label{eq:appendix_spectrum_Andrew}
\begin{align}
    \frac{a^3 n_{\chi,\threehalf,p}}{a_e^3 H_e^3} & = 
    \frac{9}{256 \pi} \, 
    \bigl( 1-r^2 \bigr)^{9/4} \, 
    r^2 \, 
    \biggl( \frac{p}{a_e m_\phi} \biggr)^{\!\!-3/2} 
    \\ & \quad \times 
    \Theta(m_\phi - m_\chi) \, 
    \Theta\bigl( \sqrt{1-r^2} \, a_\ast m_\phi < p < \sqrt{1-r^2} \, a_\RH{} m_\phi \bigr) \nn 
    \frac{a^3 n_{\chi,\onehalf,p}}{a_e^3 H_e^3} & = 
    \frac{1}{16 \pi} \, 
    \bigl( 1-r^2 \bigr)^{9/4} \, 
    r^{-2} \, 
    \biggl( 1 - \frac{3}{4} r^2 \biggr)^2 \, 
    \biggl( \frac{p}{a_e m_\phi} \biggr)^{\!\!-3/2} 
    \\ & \quad \times 
    \Theta(m_\phi - m_\chi) \, 
    \Theta\bigl( \sqrt{1-r^2} \, a_\ast m_\phi < p < \sqrt{1-r^2} \, a_\RH{} m_\phi \bigr) 
    \;.
    \nonumber
\end{align}
\end{subequations}
These expressions are equivalent to \eref{eq:a3np_Boltzmann} in the main text.  

\subsection{Sound speed not equal to unity}

In the previous appendix (A.1), we presented the Boltzmann calculation for raritron production with the implicit assumption that $|c_s| = 1$ at all times.  However, the high-mass and low-mass raritron models have a sound speed $c_s(t) \in \mathbb{C}$ that is both time-dependent and not generally equal to $1$.  In this appendix we generalize the Boltzmann calculation to allow for a nontrivial sound speed.  For simplicity we continue to assume that $c_s$ is constant in time, although we now allow it to take values $0 \leq |c_s| < 1$.  This as a rough treatment rather than a robust calculation. For example, there's no reason to expect that the rate would be controlled by a Lorentz invariant matrix element, since the system is not Lorentz invariant for $c_s \neq 1$.

We treat the effect of a constant sound speed $c_s \neq 1$ as a modification to the raritron's energy-momentum relation: $E_\chi(\kvec) = [m_{3/2}^2 + |c_s|^2 |\kvec|^2]^{1/2}$. We assume that the squared matrix element is not affected by the new physics that gives rise to $c_s \neq 1$, and we continue to evaluate $|\overline{\mathcal{M}}|^2$ using \eref{eq:msquared}.  

The delta function in \eqref{eq:Rphisq_a_steps} becomes
\begin{align}
       \delta(E_3-m_\phi) = \frac{|c_s|^{-1}m_\phi}{\sqrt{m_\phi^2-m_{3/2}^2}}\delta\left(|\kvec_3|-|c_s|^{-1}\sqrt{m_\phi^2-m_{3/2}^2}\right),
\end{align}
which yields
\begin{align}
        \int \frac{\dd^3\kvec_3}{E_3} \ \delta(E_3-m_\phi) 
    & = \frac{4\pi}{|c_s|} \frac{m_\phi}{\sqrt{m_\phi^2-m_{3/2}^2}} \int \dd|\kvec_3| \frac{|\kvec_3|^2}{\sqrt{|c_s|^2|\kvec_3|^2+m_{3/2}^2}}\ \delta\left(|\kvec_3|-|c_s|^{-1}\sqrt{m_\phi^2-m_{3/2}^2}\right) \,.
\end{align}
Expressing the physical momentum $|\kvec|$ in terms of the comoving momentum $p=a|\kvec|$,
this gives
\begin{align}
     \int \frac{\dd^3\kvec_3}{E_3} \ \delta(E_3-m_\phi) 
  & =  \frac{4\pi}{|c_s|} \frac{m_\phi}{\sqrt{m_\phi^2-m_{3/2}^2}} \frac{1}{a} \int \dd p \  \frac{ p^2}{\sqrt{|c_s|^2p^2+a^2m_{3/2}^2}}\  \delta\left(p-|c_s|^{-1}a\sqrt{m_\phi^2-m_{3/2}^2}\right) ,
\end{align}
which leads to an expression for the collision term and its derivative in terms of $a$:
\begin{align}
    C(a) & = \frac{n_\phi^2\ |\mathcal{M}|^2}{32\pi m_\phi^3} \ \frac{1}{\sqrt{1-r^2}} \frac{1}{a} \frac{1}{|c_s|} \int \dd p \  \frac{p^2}{\sqrt{|c_s|^2p^2+a^2m_{3/2}^2}} \delta\left(p-|c_s|^{-1}a\sqrt{m_\phi^2-m_{3/2}^2}\right) \, , \nonumber \\
    \frac{\dd{C}(a)}{dp} & = \frac{n_\phi^2\ |\mathcal{M}|^2}{32\pi m_\phi^4} \ \frac{1}{1-r^2} \frac{p^2}{\sqrt{|c_s|^2p^2+a^2m_{3/2}^2}} \ \ \frac{1}{a|c_s|} \ \delta\left(a-\frac{|c_s|p}{\sqrt{m_\phi^2-m_{3/2}^2}}\right)\, .
\end{align}
If $|c_s|$ is an arbitary function of $a$, the above expression is unwieldy, but some indication of the effect of having $|c_s|<1$ can be found by considering $|c_s|=\mathrm{const.}\neq1$.

Assuming $|c_s|=\mathrm{const.}$, and using $a^3 n_{\chi,p,a} = p \dd (a^3 n_{\chi,a}) / \dd p$, we find
\begin{subequations}\label{eq:nkvary}
\begin{align}
    \frac{a^3 n_{\chi,\threehalf,p}}{a_e^3 H_e^3} & = 
    \frac{9}{256 \pi} \, 
    \bigl( 1-r^2 \bigr)^{9/4} \, 
    r^2 \, 
    \biggl( \frac{|c_s|p}{a_e m_\phi} \biggr)^{\!\!-3/2} \frac{1}{|c_s|^4}
    \\ & \quad \times 
    \Theta(m_\phi - m_\chi) \, 
    \Theta\bigl( \sqrt{1-r^2} \, a_\ast m_\phi < |c_s|p < \sqrt{1-r^2} \, a_\RH{} m_\phi \bigr) \nn 
    \frac{a^3 n_{\chi,\onehalf,p}}{a_e^3 H_e^3} & = 
    \frac{1}{16 \pi} \, 
    \bigl( 1-r^2 \bigr)^{9/4} \, 
    r^{-2} \, 
    \biggl( 1 - \frac{3}{4} r^2 \biggr)^2 \, 
    \biggl( \frac{|c_s|p}{a_e m_\phi} \biggr)^{\!\!-3/2} \frac{1}{|c_s|^4}
    \\ & \quad \times 
    \Theta(m_\phi - m_\chi) \, 
    \Theta\bigl( \sqrt{1-r^2} \, a_\ast m_\phi < |c_s|p < \sqrt{1-r^2} \, a_\RH{} m_\phi \bigr) 
    \;.
    \nonumber
\end{align}
\end{subequations}
Comparing \eqref{eq:appendix_spectrum_Andrew} and\eqref{eq:nkvary}, we see that if to go from sound speed $|c_s| = 1$ to sound speed $|c_s|\neq 1$, replace $p\to|c_s|p$ and include an overall factor of $|c_s|^{-4}$.  Clearly as the sound speed drops below unity $n_p$ increases. 

\bibliographystyle{JHEP}
\bibliography{ref}

\end{document}